\documentclass[11pt]{scrartcl}

\usepackage[margin=2.0cm]{geometry}

\usepackage[utf8]{inputenc}
\usepackage[T1]{fontenc}
\usepackage[english]{babel}

\usepackage{amsmath,amssymb,amsfonts}
\usepackage{amsthm}
\usepackage{mathrsfs}
\usepackage[title]{appendix}
\usepackage{bm}
\usepackage{booktabs}
\usepackage{algorithm}
\usepackage{algorithmicx}
\usepackage{algpseudocode}
\usepackage{cancel}
\usepackage[format=plain,singlelinecheck=false,font={small,bf},labelfont=bf,labelsep=period]{caption}
\DeclareCaptionFont{xbf}{\small\bfseries\boldmath}
\usepackage{graphicx}
\usepackage{float}
\usepackage{color}
\usepackage{xcolor}
\usepackage{colortbl}
\usepackage{csquotes}
\usepackage{dcolumn}
\usepackage{fancyhdr}
\usepackage{helvet}
\usepackage{listings}
\usepackage{lscape}
\usepackage{mathpazo}
\usepackage{mathtools}
\usepackage[version=3]{mhchem}
\usepackage{microtype}
\usepackage{multirow}
\usepackage{pgfplots}
\usepackage{setspace}
\usepackage{siunitx}
\usepackage{subfigure}
\usepackage{tablefootnote}
\usepackage{txfonts}
\usepackage[normalem]{ulem}
\usepackage{xkeyval}
\usepackage{textcomp}
\usepackage{manyfoot}
\DeclareUnicodeCharacter{2212}{-}

\RedeclareSectionCommand[
  beforeskip=-3.0ex plus -1ex minus -.2ex,
  afterskip=1.5ex plus .2ex
]{section}
\RedeclareSectionCommand[
  beforeskip=-2.0ex plus -0.8ex minus -.2ex,
  afterskip=1.0ex plus .2ex
]{subsection}

\usepackage{xr}
\usepackage{hyperref}
\hypersetup{
  breaklinks=true,
  colorlinks=true,
  citecolor=blue,
  linkcolor=blue,
  filecolor=blue,
  urlcolor=blue
}
\usepackage{cleveref}

\usepackage[numbers,sort&compress]{natbib}
\setcitestyle{super}
\usepackage{natmove}

\makeatletter
\@ifundefined{abx@aux@refcontext}{%
  \newcommand\abx@aux@refcontext[2]{}%
}{}
\@ifundefined{abx@aux@sortscheme}{%
  \newcommand\abx@aux@sortscheme[1]{}%
}{}
\@ifundefined{abx@aux@cite}{%
  \newcommand\abx@aux@cite[3]{}%
}{}
\@ifundefined{abx@aux@segm}{%
  \newcommand\abx@aux@segm[5]{}%
}{}
\makeatother

\usepackage{authblk}

\usepackage{letltxmacro}
\LetLtxMacro{\oldsqrt}{\sqrt}
\renewcommand{\sqrt}[2][\mkern8mu]{\mkern-8mu\mathop{}\oldsqrt[#1]{#2}}

\def\SM{Supporting Information}

\title{Interpretable, Physics-Informed Learning Reveals Sulfur Adsorption and Poisoning Mechanisms in 13-Atom Icosahedra Nanoclusters}

\author[1]{Raiane Ferreira Monteiro}
\author[1]{Jo{\~a}o Marcos T. Palheta}
\author[1]{Tulio Gnoatto Grison}
\author[1]{Octávio Rodrigues Filho}
\author[2]{Renato Luis Tame Parreira}        
\author[3]{Diego Guedes-Sobrinho}
\author[4,*]{Celso R. C. R{\^e}go}
\author[5]{Alexandre C. Dias}
\author[6]{Krys Elly de Araújo Batista}
\author[1]{Maur{\'i}cio J. Piotrowski}

\affil[1]{Department of Physics, Federal University of Pelotas, PO Box $354$, Pelotas, RS, $96010$-$900$, Brazil}
\affil[2]{Núcleo de Pesquisas em Ciências Exatas e Tecnológicas, Universidade de Franca, Franca, SP, Brazil}
\affil[3]{Chemistry Department, Federal University of Paran\'a, 81531-980, Curitiba, Brazil}
\affil[4]{Karlsruhe Institute of Technology (KIT), Institute of Nanotechnology Hermann-von-Helmholtz-Platz, 76344, Eggenstein-Leopoldshafen, Germany}
\affil[5]{Institute of Physics and International Center of Physics, University of Bras{\'i}lia, Bras{\'i}lia $70919$-$970$, DF, Brazil}
\affil[6]{Federal Institute of Amazonas, Manaus, AM, $69086$-$475$, Brazil}

\date{} 

\begin{document}

\maketitle

\begin{abstract}
Transition-metal nanoclusters exhibit structural and electronic properties that depend on their size, often making them superior to bulk materials for heterogeneous catalysis. However, their performance can be limited by sulfur poisoning. Here, we use dispersion-corrected density functional theory (DFT) and physics-informed machine learning to map how atomic sulfur adsorbs and causes poisoning on 13-atom icosahedral clusters from 30 different transition metals (3$d$ to 5$d$). We measure which sites sulfur prefers to adsorb to, the thermodynamics and energy breakdown, changes in structure, such as bond lengths and coordination, and electronic properties, such as $\varepsilon_d$, the HOMO-LUMO gap, and charge transfer. Vibrational analysis reveals true energy minima and provides ZPE-based descriptors that reflect the lattice stiffening upon sulfur adsorption. For most metals, the metal-sulfur interaction mainly determines adsorption energy. At the same time, distortion penalties are usually moderate but can be significant for a few metals, suggesting these are more likely to restructure when sulfur is adsorbed. Using unsupervised \textit{k}-means clustering, we identify periodic trends and group metals based on their adsorption responses. Supervised regression models with leave-one-feature-out analysis identify the descriptors that best predict adsorption for new samples. Our results highlight the isoelectronic triad \ce{Ti}, \ce{Zr}, and \ce{Hf} as a balanced group that combines strong sulfur binding with minimal structural change. Additional DFT calculations for \ce{SO2} adsorption reveal strong binding and a clear tendency toward dissociation on these clusters, linking electronic states, lattice response, and poisoning strength. These findings offer data-driven guidelines for designing sulfur-tolerant nanocatalysts at the subnanometer scale.
\end{abstract}


\section{Introduction}

The challenges associated with developing, manipulating, and understanding nanoscale atomistic systems have drawn significant attention from the research community due to fast advances in nanoscience.\cite{Alivisatos1996, Schmid2011, Maity2025} In the subnanometer regime, where quantum size effects become dominant, the nanostructures made of a few tens to several hundred atoms often show properties that differ from those of their bulk crystalline counterparts.\cite{Jena2008, Castleman2009} This emergent behavior enables the materials design with tunable structural and electronic features, offering many opportunities in the energy conversion, optoelectronic, and catalytic fields.\cite{Bell2003, Aslam2018} Transition metal (TM) nanoclusters (NCs) stand out in this context because of their distinct chemical reactivity, discrete electronic states, and high surface/volume ratios.\cite{Li2003, Cao2011} The intrinsic fluxionality, finite size, and lack of translational symmetry characteristics result in a broad configurational landscape with many non-equivalent adsorption sites.\cite{Yoon2005, Sun2021, Rego_2017} As a result, TM-NCs exhibit a complex, yet highly adaptable catalytic behavior, which is particularly important in real-world scenarios (realistic environmental conditions) where toxic poisoning species and/or competing reaction pathways may be present.\cite{Campbell2013}

The catalytic deactivation caused by \ce{S}-containing species is a long-standing dilemma in heterogeneous catalysis.\cite{Bartholomew2001} This challenge arises because sulfur atoms and sulfur-bearing molecules, such as \ce{H2S}, \ce{SO2}, and \ce{SOx}, show strong affinity for metal active sites, frequently leading to partial or complete catalytic activity loss.\cite{Mesilov2021, Liu2019} Although adsorption mechanisms on extended surfaces are reasonably well disseminated/understood, the behavior of \ce{S} adsorption on subnanometer clusters remains far less explored. This knowledge gap is mainly due to pronounced finite-size effects, enhanced fluxionality, and the low coordination of surface atoms in TM-NCs.\cite{Fielicke2004,Heiz2007} Consequently, developing robust nanocatalysts requires a fundamental understanding of \ce{S}--NC interactions at the atomistic scale.\cite{Norskov2011}

In TM-NCs, reactivity is governed by the electronic structure, which deviates from that of bulk metals due to quantum confinement and discrete energy levels.\cite{Hakkinen2012} In this line, systematic studies have shown that adsorption energies, preferred adsorption sites, and even reaction mechanisms can vary significantly across isomers separated by only a few tenths of an electronvolt.\cite{Ferrando2008} For specific geometrical motifs, such as the icosahedron (ICO), for example, the interplay between geometric shell closure and $d$-band occupation creates interesting periodic trends in adsorption behavior across the TM series.\cite{Peraca2020, Piotrowski_13172_2024} Thus, understanding how \ce{S} interacts with these electronically diverse motifs provides valuable insights into the design of more resilient catalytic nanostructures.

Previous experimental and theoretical studies have pointed out the critical impact of \ce{S} poisoning on metal nanoparticles and NCs.\cite{Bartholomew2001} For instance, studies on \ce{Au}, \ce{Pd}, and \ce{Pt} NCs have demonstrated that \ce{S} can induce substantial structural rearrangements, change the $d$-band center, and even lead to NC fragmentation at elevated coverages.\cite{Hakkinen2012, Say2022} On the other hand, another work on subnanometer clusters shows that \ce{S} adsorption may stabilize unexpected geometries.\cite{Yin2021} In addition to recent machine learning (ML) advances, which have accelerated materials discovery, enabling large-scale exploration of NC potential-energy landscapes,\cite{Deringer2019, Zuo2020, Chmiela2018} the ML models trained on first-principles (density functional theory-DFT) data have been effective in identifying periodic trends, predicting adsorption energies, and capturing subtle structure, and property relationships in metal NCs.\cite{Xie2018} By incorporating these approaches (DFT+ML) into our calculations and analysis, we not only accelerate the screening of \ce{TM}--\ce{S} interactions but also uncover systematic behaviors that would be difficult to detect through DFT alone, further strengthening the predictive power of nanocatalyst design.

Here, we investigate sulfur adsorption and incipient poisoning on \num{13}-atom icosahedral transition-metal nanoclusters (\ce{TM13}) spanning the \num{3}$d$, \num{4}$d$, and \num{5}$d$ series. The icosahedral (ICO) motif is among the most stable geometries in the subnanometer regime,\cite{Baletto2005, Johnston1998, Imaoka2013, Piotrowski_155446_2010}, consisting of a central atom surrounded by 12 surface atoms at the vertices of a regular ICO, and providing a high-symmetry, close-packed platform with well-defined adsorption sites (top, bridge, hollow). Using dispersion-corrected, spin-polarized DFT, we characterize pristine and \ce{S}-adsorbed ICO clusters through a set of physically motivated descriptors that span energetics (e.g., $E_\text{ads}$ and its interaction/distortion decomposition), electronic structure (e.g., $\varepsilon_d$, $E_\text{gap}$, charge transfer), and lattice dynamics (vibrational fingerprints and ZPE-based descriptors). We then introduce a physics-informed interpretation of sulfur adsorption and poisoning trends by combining these descriptors with interpretable machine learning (ML): unsupervised clustering maps systematic, periodic-table-level similarities in adsorption response, while supervised, model-agnostic LOFO analyses quantify which descriptors contribute robustly to out-of-sample prediction and thus to mechanistic interpretation. This integrated DFT+ML framework yields transferable rules linking adsorption strength to electronic-state alignment and adsorption-induced lattice response. It motivates the selection of the isoelectronic triad \ce{Ti13}, \ce{Zr13}, and \ce{Hf13} as representative, chemically resilient case studies for explicit \ce{SO2} adsorption, where follow-up DFT provides direct validation of the ML-guided trends.

\section{Methodology and Computational Details}

\paragraph*{Density Functional Theory:} Our calculations were performed using spin-polarized DFT\cite{Hohenberg_B864_1964,Kohn_A1133_1965} as implemented in the Vienna \textit{Ab Initio} Simulation Package (VASP),\cite{Kresse_13115_1993,Kresse_11169_1996} employing the projector augmented-wave (PAW) method\cite{Blochl_17953_1994,Kresse_1758_1999} to describe the electron–ion interactions. The exchange-correlation energy was described by the generalized gradient approximation (GGA) with the Perdew–Burke–Ernzerhof (PBE) functional,\cite{Perdew_3865_1996, Perdew_1396_1997} including dispersion corrections via the empirical D3 corrections, as proposed by Grimme,\cite{Grimme_154104_2010, Grimme_5105_2016, Rego_2015} to include the attractive non-local long-range van der Waals (vdW) interactions. In addition, spin–orbit coupling (SOC) was included and tested, which could be particularly important for \num{4}$d$ and \num{5}$d$ elements, where relativistic effects may be significant. Otherwise, in our approach, the valence electrons were treated using a scalar-relativistic approximation, while the core electrons were treated fully relativistically.\cite{Koelling_3107_1977, Takeda_43_1978}

The TM-NC systems (for pristine and atomically/molecular adsorbed cases) were modeled from an isolated \num{13}-atom ICO, i.e., as non-periodic systems through a cubic box (supercell) with a side length of \SI{20}{\angstrom}, providing a minimum separation distance of approximately \SI{12}{\angstrom}, which is enough to avoid the spurious self-interactions among the system and their periodic images. All systems were fully optimized until residual forces were below \SI{0.01}{\electronvolt\per\angstrom}, using a plane-wave cutoff energy of \SI{500}{\electronvolt} and an energy convergence threshold of \SI{E-6}{\electronvolt} in the self-consistent cycle. For density of states (DOS), we used a cutoff energy value that was \num{1.5} times higher. For Brillouin zone integration, a single $\textbf{k}$-point, specifically, the $\Gamma$-point, was used for all calculations. A small Gaussian smearing parameter of \SI{1}{\milli\electronvolt} was applied to prevent fractional occupation of electronic states.  

It is important to mention that the energetic numerical precision near the equilibrium geometry calculated with the VASP package is on the order of \SI{0.001}{\electronvolt} per atom.\cite{Lejaeghere_1415_2016} Furthermore, the PBE functional has no fitted parameters, which represents an advantage in studying NCs, since for these systems there is reduced experimental data. The PBE accuracy is \SI{0.28}{\electronvolt} when measured relative to a known isomerization-energy dataset, considered difficult for theory.\cite{Mardirossian_2315_2017}. In our case, we expect better accuracy, since we are comparing relative energies within families of related NC species. Finally, the vibrational frequency ($\nu$) calculations were performed to determine the $\num{3}N - \num{6}$ vibrational modes for all systems, for which the Hessian matrix was calculated using finite differences, as implemented in VASP, considering the displacing of each atom in each direction by $\pm$\SI{0.01}{\angstrom}. Further computational details, including convergence tests, are provided in the \SM (Tables S1-S4). 

In the last part of this work, after the selection process, to obtain the most stable \ce{SO2} adsorption site on \ce{Ti13}, \ce{Zr13}, and \ce{Hf13} NCs, we explored the analyte-substrate potential energy surface employing a strategy based on \textit{Ab{~}Initio} Molecular Dynamics (AIMD) simulations and additional initial configurations built by design-principles (to complement the AIMD calculations). The AIMD simulations started with \ce{SO2} and NC atomic configurations in arbitrary orientations, followed by a thermalization process at \SI{300}{\kelvin}, using a Nos{\'e}-Hoover thermostat within an NVT ensemble.\cite{Nose_255_1984, Hoover_1695_1985} Our AIMD simulations were run for \SI{5}{\pico\second}, with a time step of \SI{1}{\femto\second}, to generate the final adsorbed snapshot, which was subsequently structurally optimized through DFT-PBE+D3 calculations. 

Our simulation workflow follows the FAIR (Findable, Accessible, Interoperable, Reusable) and TRUE (Transparent, Reproducible, Usable by Others, and Extensible) data principles.\cite{Rego_2022, Schaarschmidt2021, Bastos_2025}. Here, we ensure that the complete set of data and methods produced in this study is open, reproducible, and straightforward for others to reuse or build upon. In line with these guidelines, we provide full provenance of data sources, document uncertainties where applicable, and include clear reuse instructions.\cite{Rego_2025, deAraujo2024, Soleymanibrojeni_2024, Bekemeier_2025, Dalmedico_2024}. All DFT inputs and outputs, processed descriptor tables, machine-learning workflows, trained models, and analysis scripts generated in this work are openly available in the public repository \url{https://github.com/KIT-Workflows/Physics-Informed-Nanoclusters}, enabling direct reuse, validation, and extension by the community.

\begin{figure*}[htb!]
\centering
\includegraphics[width=1.0\linewidth]{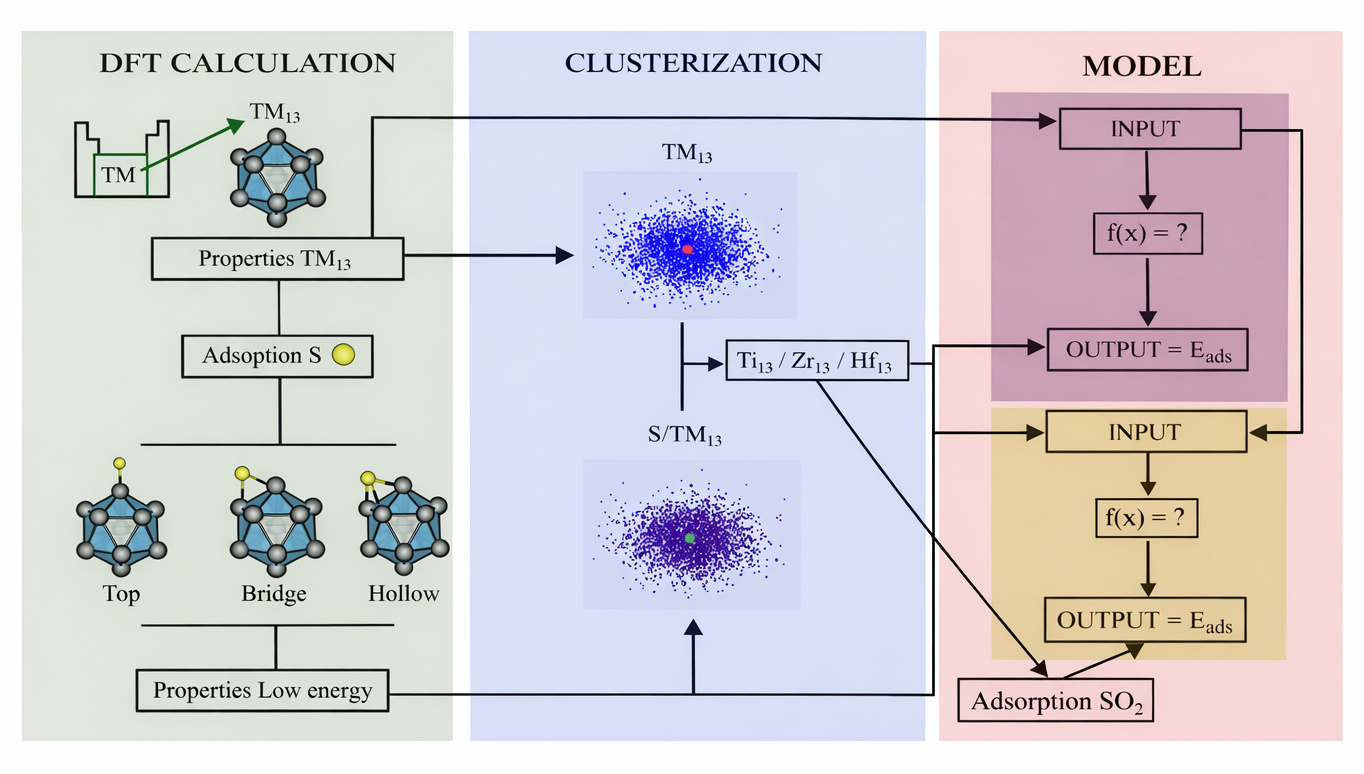}
\caption{Schematic workflow illustrating the combined first-principles and machine-learning strategy adopted in this work. Starting from DFT calculations, \num{13}-atom ICO TM-NCs in pristine form (\ce{TM13}) and after atomic \ce{S} adsorption (\ce{S{\text{/}}TM$_{13}$}), considering top, bridge, and hollow binding sites. Structural, energetic, electronic, and vibrational descriptors extracted from DFT calculations were subsequently standardized and used as input for unsupervised clustering analyses using \textit{k}-means and principal component analysis (PCA), enabling the identification of chemically similar groups across the TM series. Based on the clustering outcomes, selected nanoclusters (\ce{Ti13}, \ce{Zr13}, and \ce{Hf13}) were further employed in explicit \ce{SO2} adsorption calculations. Finally, the descriptor set was used to train the ElasticNet regression models, allowing the prediction and rationalization of \ce{S} adsorption energetics to provide a selection of chemically resilient NCs.}
\label{fig1}
\end{figure*}

\paragraph*{Atomic Configurations:} We have considered the following systems: \num{13}-atom ICO NCs of all TM elements in ($i$) pristine case (bare clusters, \ce{TM13}), in first moment, and ($ii$) atomic-sulfur-adsorbed NC case (\ce{S{\text{/}}TM$_{13}$}), in second moment; after that, TM selected \num{13}-atom ICO NCs in ($iii$) molecular-sulfur-based-adsorbed NC case (\ce{SO$_2${\text{/}}TM$_{13}$}), as illustrated schematically in Figure~\ref{fig1}. As well as the respective free atom calculations of all these systems and the isolated \ce{SO2} molecule calculation. Therefore, it is evident that our main focus is to test the composition (among all TM elements) rather than geometric motif; consequently, the structural configuration considered for our \num{13}-atom NCs of all TM elements are initially set to the ICO, which is the lowest energy configuration for several \ce{TM13} NCs.\cite{Piotrowski_155446_2010, Chaves_15484_2017} The ICO motif is a quasi-spherical particle, with close-packed high-symmetry (I$_h$), composed by a central atom and \num{12} equidistant atoms. All ICO-based calculations performed here had their geometries relaxed, without any symmetry or spin constraints, to allow structural and magnetic symmetry breaking and a more complete exploration of the potential energy surface.

The two main sets of configurations: \ce{TM13} and \ce{S{\text{/}}TM$_{13}$} are showed in \SM (Figure S1). The first set (\ce{TM13}) is used as a substrate for the adsorption of \ce{S} atoms. We explored all non-equivalent positions, considering an approach similar to that used for low-Miller-index surfaces, i.e., onefold (top), twofold (bridge), and threefold (hollow) sites on NCs. After a relative total-energy analysis of the adsorption sites, we obtained the \ce{S{\text{/}}TM$_{13}$} set. Consequently, we systematically considered all non-equivalent sites with distinct coordination (with respect to the ICO motif) and compositions (with respect to all TM elements from the periodic table). Then, after our selection procedure, we have chosen three TM-NCs for the last calculation step, i.e., \ce{SO$_2${\text{/}}TM$_{13}$}. For this case, exploring the surface potential energy through AIMD simulations and placing the \ce{SO2} molecule in all non-equivalent positions on NCs, considering trivial sites and specific cases with side-on and end-on orientations. It is important to note that AIMD simulations provide a more complete set of adsorption configurations and allow testing of their thermal stability.

\paragraph*{Property Analysis:} To characterize the energetic stability of the ICO NCs, we calculated the binding energy per atom, $E_{\text{b}}^{\ce{TM13}}$, as
\begin{equation}
E_{\text{b}}^{\ce{TM13}} = \frac{E^{\ce{TM13}}_{\text{tot}} - 13 E^{\ce{TM}~\text{free-atom}}_{\text{tot}}}{13}~,
\end{equation}
where $E^{\ce{TM13}}_{\text{tot}}$ is the total energy of the NC and $E^{\ce{TM}~\text{free-atom}}_{\text{tot}}$ the energy of each isolated atomic species. This metric enables comparison of relative stabilities across different TM-NCs. In analogy, to characterize the energetic stability of the \ce{S{\text{/}}TM$_{13}$} systems, we calculated the binding energy per atom, $E_{\text{b}}^{\ce{S{\text{/}}TM13}}$, as
\begin{equation}
E_{\text{b}}^{\ce{S{\text{/}}TM13}} = \frac{E^{\ce{S{\text{/}}TM13}}_{\text{tot}} - E^{\ce{S}~\text{free-atom}}_{\text{tot}} - 13 E^{\ce{TM}~\text{free-atom}}_{\text{tot}}}{14}~,
\end{equation}
where $E^{\ce{S{\text{/}}TM13}}_{\text{tot}}$ is the total energy of the \ce{S{\text{/}}TM$_{13}$} and $E^{\ce{S}~\text{free-atom}}_{\text{tot}}$ the energy of sulphur atomic species.

The adsorption energy ($E_\text{ads}$) of an atomic or molecular species ($X$ = \ce{S} or \ce{SO2}) was defined as
\begin{equation}
E_\text{ads} = E_\text{tot}^{X{\text{/}}\ce{TM13}} - E_\text{tot}^{\ce{TM13}} - E_\text{tot}^X~,
\end{equation}
where $E_\text{tot}^{X{\text{/}}\ce{TM13}}$, $E_\text{tot}^{\ce{TM13}}$, and $E_\text{tot}^{X}$ are the total energies of the adsorbed system, pristine NC, and isolated adsorbate, respectively. To gain further insight into the adsorption mechanism, $E_\text{ads}$ was decomposed into interaction and distortion contributions:\cite{Yonezawa_4805_2021,Sousa_139945_2022,Felix_1040_2023}
\begin{equation}
E_\text{ads} = \Delta E_\text{int} + 13\Delta E_\text{dis}^{\ce{TM13}} + \Delta E_\text{dis}^{X}~,
\end{equation}
where $\Delta E_\text{int}$ is the interaction energy between frozen fragments, disregarding the energetic contribution from structural distortions, which is given by
\begin{equation}
\Delta E_{\text{int}} = E_\text{tot}^{X{\text{/}}\ce{TM13}} - E_\text{tot}^{\ce{TM13}~\text{frozen}} - E_\text{tot}^{X~\text{frozen}}~,
\end{equation} 
while $\Delta E_\text{dis}^{\ce{TM13}}$ per atom and $\Delta E_\text{dis}^{X}$ per adsorbate denote the distortion penalties of the NC and adsorbate, respectively, given by
\begin{equation}
\Delta E_{\text{dis}}^{\ce{TM13}} =  \frac{E_\text{tot}^{\ce{TM13}~\text{frozen}} - E_\text{tot}^{\ce{TM13}}}{13}~,
\end{equation} 
and
\begin{equation}
\Delta E_{\text{dis}}^{X} = E_\text{tot}^{X~\text{frozen}} - E_\text{tot}^{X}~, 
\end{equation} 
where $E_\text{tot}^{\ce{TM13}~\text{frozen}}$ and $E_\text{tot}^{X~\text{frozen}}$ are the total energies of the frozen NC and $X$ at their original positions within the \ce{X{\text{/}}TM13} system without $X$ and \ce{TM13} parts, respectively. These values indicate the energy required to distort configurations from their initial to adsorbed stages; consequently, $\Delta E_\text{dis}^{X}$ term is null for $X$ = \ce{S}.   

We have obtained the ($\num{3}N - \num{6}$) vibrational frequencies ($\nu$) by calculating the dynamic Hessian matrix elements within the finite difference method implemented in VASP. This approach involved two atomic displacements, with each atom moved in each direction (positive and negative) by \SI{0.01}{\angstrom}, allowing the numerical construction of second derivatives of the total energy with respect to atomic displacements. The resulting Hessian matrix was subsequently diagonalized to yield the vibrational eigenmodes and their corresponding frequencies. The vibrational analysis has multiple purposes. ($i$) The absence of imaginary frequencies confirms that all optimized pristine and \ce{S}- or \ce{SO2}-adsorbed NCs correspond to true local minima on the potential energy surface. ($ii$) The complete vibrational spectra provide a characteristic fingerprint for each system, enabling detailed comparisons of structural motifs and adsorption-induced changes. ($iii$) In the specific case of the isolated \ce{SO2} molecule, the computed vibrational frequencies were validated against available experimental and theoretical data, ensuring the reliability of the adopted computational protocol. These reference values also allow direct comparison between gas-phase and adsorbed \ce{SO2}, helping elucidate frequency shifts associated with charge transfer, bond weakening or strengthening, and/or symmetry breaking upon adsorption on the TM-NCs. ($iv$) Finally, the minimum and maximum vibrational frequencies, together with the zero-point energy (ZPE), were systematically extracted for all systems. The ZPE was computed as
\begin{equation}
E_{\mathrm{ZPE}} = \frac{1}{2}\sum_i h \nu_i~,
\end{equation}
where the sum runs over all vibrational modes. These vibrational descriptors were subsequently incorporated into the ML model as physically motivated features, complementing structural, electronic, and energetic inputs.

Structural descriptors included the average bond length, $d_{\text{av}}$, and effective coordination number, ECN, obtained from the effective coordination concept.\cite{Hoppe_25_1970, Hoppe_23_1979}. In addition to calculating $d_{\text{av}}$ and ECN for pristine systems, to quantify structural changes induced by adsorption, we also computed relative deviations as
\begin{equation}
{\Delta}d_\text{av} = \frac{(d_\text{av,ads} - d_\text{av}) \times 100}{d_\text{av}}~,
\end{equation}
and
\begin{equation}
{\Delta}\text{ECN} = \frac{(\text{ECN}_\text{ads} - \text{ECN}) \times 100}{\text{ECN}}~,
\end{equation}
where the values of $d_{\text{av, ads}}$ and ECN$_{\text{ads}}$ are obtained after adsorption, with molecules subsequently removed from the analyses. Thus, ${\Delta}d_\text{av}$ and ${\Delta}\text{ECN}$ parameters provide a direct measure of NC expansion/contraction, distortions, and coordination changes upon adsorption. We also considered the minimum NC--adsorbate distance, i.e., $d_{\ce{TM}-S}$, using VESTA software.\cite{Momma_653_2008}

Electronic structure analyses included projected DOS (PDOS) calculations and charge redistribution analysis. From PDOS, we have obtained the center of gravity of the occupied $d$-states, $\varepsilon_{\text{d}}$, which is derived from the $d$-band model.\cite{Hammer_71_2000} It provides the energetic position of the metal $d$-states relative to the Fermi level, and is related to the adsorption energy magnitude of an adsorbate on the \ce{TM} systems. Consequently, a $\varepsilon_{\text{d}}$ closer to the Fermi energy (\SI{0.0}{\electronvolt}), implies stronger hybridization between metal $d$-states and adsorbate orbitals, leading to enhanced adsorption strength. Additionally, we evaluated the HOMO--LUMO energy gap ($E_{\text{gap}}$) of the NCs, defined as the energy difference between the highest occupied molecular orbital (HOMO) and the lowest unoccupied molecular orbital (LUMO). It is a direct measure of the electronic hardness and chemical reactivity of finite systems, with smaller gaps typically associated with higher polarizability, enhanced charge-transfer capability, and increased chemical activity.

The charge findings, including charge redistribution effects, were evaluated by Bader charge analysis,\cite{Bader_1994, Tang_084204_2009} in which the total electron density is partitioned into Bader volumes (atomic basins) defined by zero-flux surfaces of the charge-density gradient, $\nabla n(\mathbf{r})$. The net charge on atom $\alpha$ is given by
\begin{equation}
Q_{\alpha}^{\text{Bader}} = Z_{\alpha} - \int_{V_{\alpha}} n(\mathbf{r}) \, d^{3}r~,
\end{equation}
where $Z_{\alpha}$ is the valence charge of atom $\alpha$ and $V_{\alpha}$ its corresponding Bader volume. Together, these energetic, vibrational, structural, and electronic descriptors capture the interplay between electronic stability, reactivity, and adsorption strength, enabling a comprehensive analysis of the interaction mechanisms between TM-NCs and \ce{S}-containing adsorbates.

\paragraph*{Machine Learning Models:}
To gain more insight about our system, we are applying a set of Machine-learning (ML) techniques to (i) rationalize periodic trends in atomic \ce{S} adsorption across the set of \num{30} \ce{TM13} ICO nanoclusters and (ii) guide the selection of representative systems for subsequent explicit \ce{SO2} adsorption studies. Given the limited dataset size, our ML strategy prioritizes physical interpretability, robustness under cross-validation, and consistency with first-principles descriptors, rather than maximizing raw predictive performance.

We derived a set of descriptors from DFT calculations to describe both pristine \ce{TM13} nanoclusters and their \ce{S}-adsorbed forms (\ce{S/TM13}). These features include energetic, structural, vibrational, and electronic properties related to adsorption and nanocluster stability. Examples are per-atom binding energy \(E_{\mathrm{b}}\), structural measures like \(d_{\mathrm{av}}\) and ECN, vibrational values such as \(\nu_{\min}\), \(\nu_{\max}\), and \(E_{\mathrm{ZPE}}\), and electronic descriptors like \(\varepsilon_{d}\), Bader charge, and \(E_{\mathrm{gap}}\). We also included adsorption-related energies (\(E_{\mathrm{ads}}\), \(\Delta E_{\mathrm{int}}\), \(\Delta E_{\mathrm{dis}}\)) when relevant. Before applying machine learning, we standardized all numerical descriptors using the \texttt{StandardScaler} from \textsc{Scikit-learn}\cite{Pedregosa2011} so that features with different units would contribute equally in both distance-based and regularized regression analyses.

First, we used an unsupervised learning classification approach to find similarities and groupings among the \ce{TM13} nanoclusters in descriptor space. Clustering was performed separately for pristine \ce{TM13} and \ce{S/TM13} systems using the \textit{k}-means algorithm in \textsc{Scikit-learn} \cite{Pedregosa2011}. We set the number of clusters to \(k=10\), based on the idea that small groups of chemically similar elements would form compact regions in descriptor space across the \num{30} transition metals. We then compared the clustering results for pristine and \ce{S}-adsorbed systems (Figures~\ref{fig1} and~\ref{fig4}) to identify elements that remained close in both cases, suggesting they exhibit similar energetic, structural, vibrational, and electronic responses to \ce{S} adsorption. This analysis identified \ce{Ti13}, \ce{Zr13}, and \ce{Hf13} as a clear, physically meaningful group that shows strong \ce{S} binding and moderate changes upon adsorption, so we chose them for detailed \ce{SO2} adsorption calculations.

Second, to improve model interpretability, we trained a set of supervised regression models to link DFT-derived descriptors with adsorption energetics. Here, we focus on predicting atomic-sulfur adsorption energies on \ce{TM13} and identifying which pristine descriptors provide a generalizable signal. To balance robustness and interpretability, we used three complementary regressors: (i) \texttt{ElasticNetCV}\cite{Zou2005, Pedregosa2011}, which combines L1 and L2 regularization and selects among correlated descriptors; (ii) \texttt{RidgeCV}, which uses L2 regularization to spread weight across collinear features; and (iii) the Explainable Boosting Machine (EBM)\cite{nori2019_interpretml, Lou2013}, an interpretable boosted generalized additive model that captures nonlinear main effects while staying transparent. We evaluated models and quantified feature utility using group cross-validation to reduce optimistic bias and test generalization across chemically distinct subsets instead of within a single correlated series.

To better understand how collinearity affects our models, we used a leave-one-feature-out (LOFO) protocol (Figure~\ref{fig5}). For each feature \(f\), we recalculated cross-validated predictions after removing \(f\) and measured how much generalization changed. We report \(\Delta R^{2}(f)=R^{2}_{\mathrm{full}}-R^{2}_{-f}\), \(\Delta\mathrm{MAE}(f)=\mathrm{MAE}_{-f}-\mathrm{MAE}_{\mathrm{full}}\), and \(\Delta\mathrm{RMSE}(f)=\mathrm{RMSE}_{-f}-\mathrm{RMSE}_{\mathrm{full}}\). Here, "full" means the baseline model with all features, and "-f" means the model with feature \(f\) removed. Positive \(\Delta\) values indicate that the feature improves generalization, since removing it worsens performance. Negative \(\Delta\) values suggest the feature is redundant or adds noise when descriptors are correlated. When LOFO results are similar across ElasticNetCV, RidgeCV, and EBM, it gives a reliable, model-independent view of which pristine descriptors matter most for \ce{S} adsorption trends. This also helps us interpret sulfur adsorption and poisoning as effects of both electronic stability/reactivity and lattice-dynamical changes.

The physic-informed model analysis is used as a rationalization and selection layer rather than a replacement for first-principles calculations: it identifies descriptors with robust explanatory value, supports the clustering-guided choice of representative systems (\ce{Ti13}, \ce{Zr13}, \ce{Hf13}), and provides an interpretable bridge between periodic DFT trends and the subsequent explicit \ce{SO2} adsorption results.

\section{Results and Discussion}

All findings presented here originated from PBE-D3 protocol optimizations, which resulted from protocol tests considering only PBE plain calculations and hybrid protocols with vdW D3 and/or SOC corrections for all \num{30} TM-NCs and \ce{S{\text{/}}TM$_{13}$} systems. For illustrative purposes, some of these protocol tests are presented in \SM (Figures S2 and S3).

\subsection{Pristine Nanoclusters}

\begin{figure*}[htb!]
\centering
\includegraphics[width=0.99\linewidth]{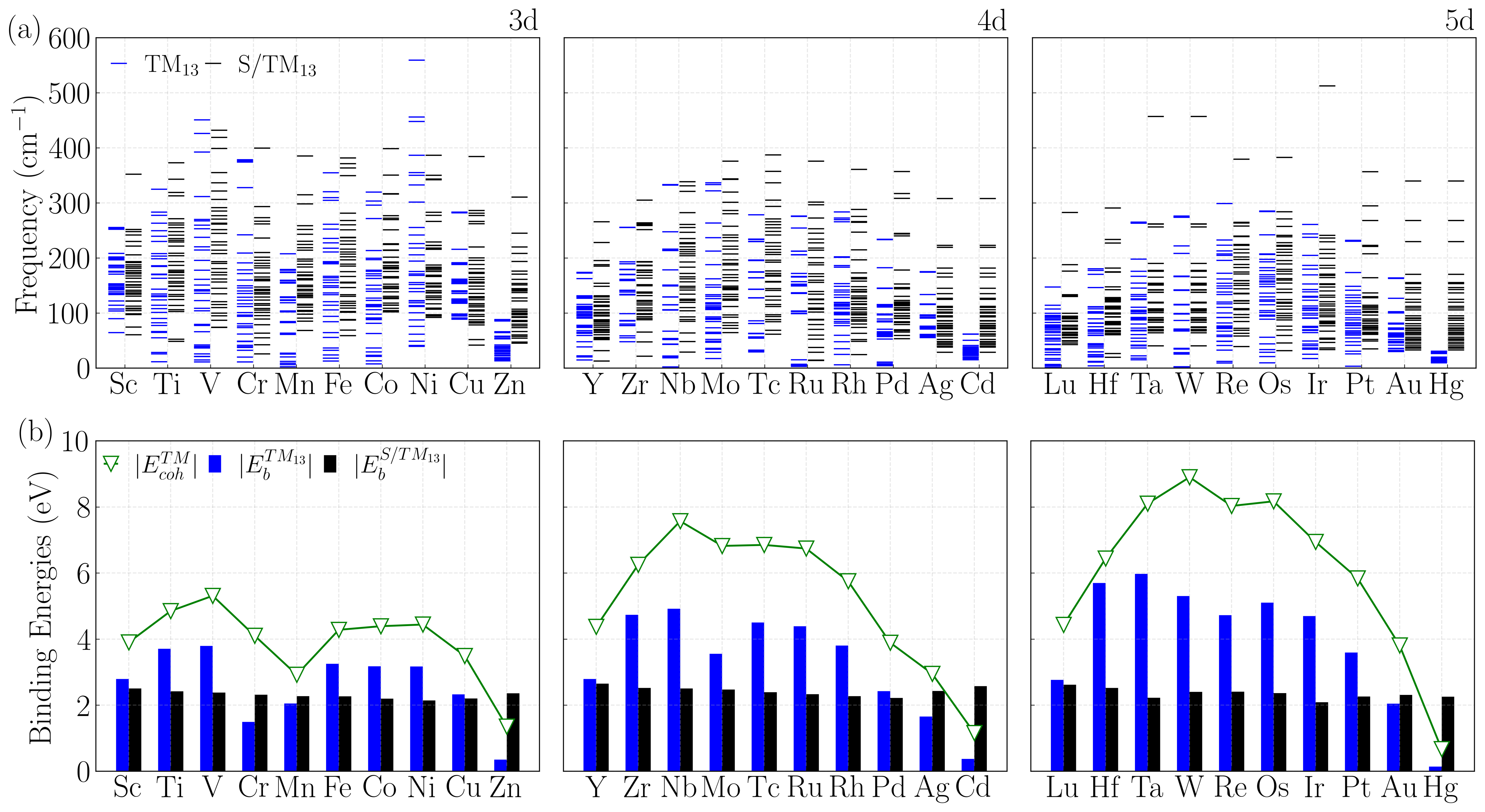}
\caption{Vibrational and energetic trends of \ce{TM13} and \ce{S{\text{/}}TM$_{13}$} across the \num{3}$d$--\num{5}$d$ series. (a) Vibrational frequencies for pristine \ce{TM13} (blue) and \ce{S{\text{/}}TM$_{13}$} (black) systems. (b) Per-atom $E_{\text{b}}^{\ce{TM13}}$ (blue) and $E_{\text{b}}^{\ce{S{\text{/}}TM13}}$ (black), compared with bulk cohesive energies ($E_{\text{coh}}^{\ce{TM}}$, green).}
\label{fig2}
\end{figure*}

After PBE+D3 optimizations, the NCs consistently preserved the expected geometry, highlighting the robustness of the ICO motif at the subnanometer scale (see Figure S1 in \SM). Vibrational frequency analysis confirmed the dynamical stability of all structures, with no imaginary modes detected, as shown in Figure~\ref{fig2}(a). In addition, the vibrational frequencies decrease systematically from \num{3}$d$ to \num{5}$d$ elements. This trend reflects the interplay between atomic mass and $d$-orbital characteristics: lighter \num{3}$d$ elements with compact $d$-orbitals form shorter, stiffer bonds, while heavier \num{5}$d$ metals experience relativistic $s$-contraction and $d$-expansion, leading to weaker bonding.

The binding energies per atom, Figure~\ref{fig2}(b), follow an approximately parabolic dependence on atomic number within each $d$-series. Maximum stability is reached for mid-series elements, consistent with optimal occupation of bonding states and minimal occupation of antibonding states, as described by the cluster orbital model.\cite{Pettifor1995} Important to note a reasonably good agreement between $E_{\text{b}}^{\ce{TM13}}$ and the cohesive energy for \ce{TM} bulk systems ($E_{\text{coh}}^{\ce{TM}}$), although finite-size and surface effects produce systematic offsets (higher per-atom binding for extended solids in some species and enhanced NC stabilization in others). Additionally, deviations in stability for certain elements, e.g., \ce{Cr} with its half-filled $d^5$ configuration, highlight the role of magnetism and local electronic effects. 

The results for $d_\text{av}$ and ECN (shown in Figures S2 and S3 in \SM) corroborate these periodic trends. While $d_\text{av}$ reflects the balance between bonding/antibonding occupation, the ECN values remain close to the ideal ICO reference (\num{6.46}), indicating that overall NC geometry is preserved even as bond lengths and binding strengths vary. Together, these descriptors establish a coherent picture: pristine ICO NCs are stable across the periodic table, with periodic modulations driven primarily by $d$-orbital filling.

\subsection{Sulfur Adsorption on Nanoclusters}

The \ce{S} adsorption on \ce{TM13} NCs was systematically investigated by considering the three basic adsorption sites on ICO surface, i.e., top, bridge, and hollow. For each TM, full structural relaxations were performed, leading to the optimized \ce{S{\text{/}}TM$_{13}$} configurations, which are presented in the \SM (Figure S1). For most cases, the hollow site is energetically preferred, reflecting the enhanced coordination of \ce{S} and the more efficient charge redistribution enabled by multi-center bonding. Nevertheless, notable exceptions are identified: for \ce{W} NCs the bridge site becomes favored, while for \ce{Ir} the top site is stabilized. These systems exhibit comparatively large distortion energies, indicating a pronounced geometrical resistance of the ICO cage to \ce{S} binding. Such cases, together with selected hollow-site adsorptions accompanied by significant deformation, correlate with anomalous \ce{S}--metal bond lengths, emphasizing that local geometry and electronic structure may override general coordination-based preferences.

Figure~\ref{fig2} illustrates that \ce{S} adsorption induces systematic changes in both the energetic and vibrational responses of the ICO \ce{TM13} NCs, with trends that depend sensitively on the TM species. As shown in panel (a), the vibrational spectra of the \ce{S{\text{/}}TM$_{13}$} systems differ markedly from those of the pristine clusters, displaying additional high-frequency modes and a broader frequency distribution. Concurrently, panel (b) reveals a clear modification of the binding-energy behavior, i.e., the pronounced parabolic dependence of the per-atom $E_{\text{b}}$ observed for pristine NCs is significantly flattened after \ce{S} adsorption. The emergence of high-frequency vibrational features is consistent with the formation of localized and relatively stiff \ce{TM}--\ce{S} bonds, which introduce modes with larger effective force constants despite the increased \ce{S} mass. Mode-projection analysis confirms that these contributions are predominantly localized at the adsorption site. The attenuated dependence of the $E_{\text{b}}$ on atomic number suggests two complementary effects, the stabilizing contribution of the \ce{TM}--\ce{S} interaction, which is of comparable magnitude across several TMs, and a partial compensation of the intrinsic variations in intra-NC bonding that dominate the energetics of pristine \ce{TM13} NCs.

Deviations from these general trends are observed for elements with half-filled or fully filled $d$-shells, such as \ce{Cr}, \ce{Mn}, and \ce{Zn}, as well as some heavier congeners (\ce{Ag}, \ce{Cd}, \ce{Au}, and \ce{Hg}). These cases highlight the role of magnetism and local electronic structure in determining adsorption energetics. In several systems, \ce{S} adsorption modifies the local magnetic moments and orbital occupations, thereby altering the balance between bonding and antibonding states and ultimately affecting the binding strength. Overall, the combined energetic and vibrational descriptors indicate that \ce{S} adsorption reorganizes local bonding more strongly than would be expected from bulk cohesive trends alone. Consequently, reliable predictions of \ce{S{\text{/}}TM$_{13}$} stability require explicit consideration of adsorbate--NC electronic interactions and adsorption-induced geometric relaxation.

\begin{figure*}[htb!]
\centering
\includegraphics[width=1.0\linewidth]{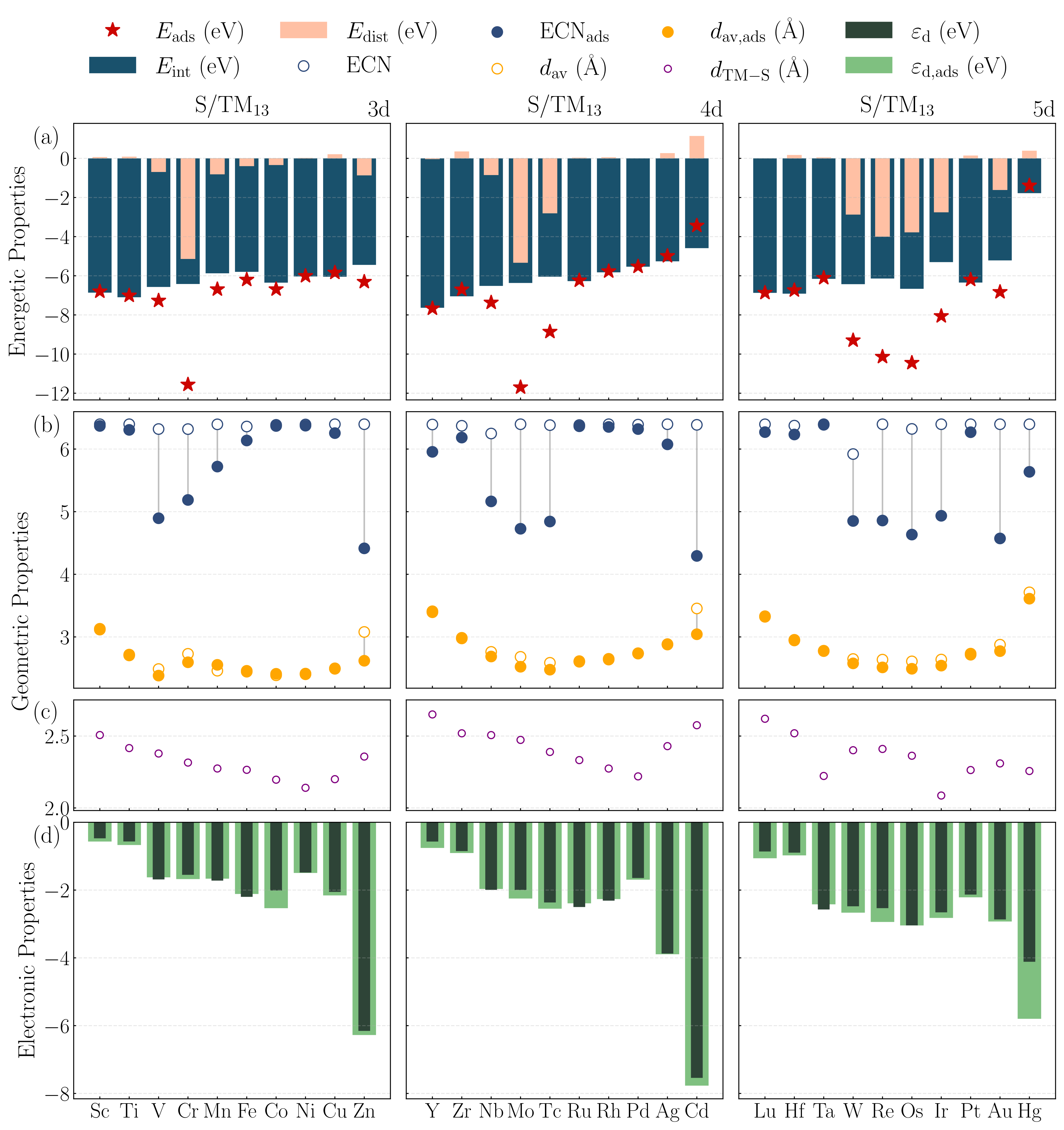}
\caption{Energetic, structural, and electronic descriptors for \ce{S} adsorption on \ce{TM13} NCs across the \num{3}$d$, \num{4}$d$, and \num{5}$d$ series. (a) Adsorption energy decomposition into interaction ($\Delta E_\text{int}$) and distortion ($\Delta E_\text{dis}^{\ce{TM13}}$) contributions, together with total adsorption energies ($E_\text{ads}$). (b) Effective coordination number (ECN) and average \ce{TM}-\ce{TM} bond length ($d_\text{av}$) before and after \ce{S} adsorption. (c) \ce{TM}-\ce{S} bond distance ($d_{\ce{TM}-\ce{S}}$) for each \ce{TM}. (d) Center of gravity of the occupied $d$-states ($\varepsilon_{\text{d}}$) for pristine and \ce{S}-adsorbed NCs.}
\label{fig3}
\end{figure*}

The $E_\text{ads}$ values reported in Figure~\ref{fig3}(a) are negative for all systems, confirming the thermodynamic favorability of \ce{S} binding on \ce{TM13} NCs. The $E_\text{ads}$ magnitude spans from \SI{1.39}{\electronvolt} for \ce{Hg} to \SI{11.70}{\electronvolt} for \ce{Mo}, with most values lying in the \num{4}--\SI{10}{\electronvolt} range. A clear strengthening of adsorption is observed when moving from the \num{3}$d$ to the \num{5}$d$ series, which is consistent with the increased radial extent of the $d$ orbitals and the growing importance of scalar-relativistic effects in heavier elements. Both factors enhance $d$--$p$ overlap with \ce{S}, thereby increasing bonding strength. While strong \ce{S} binding may facilitate adsorbate activation, it also hampers desorption, in line with the Sabatier principle,\cite{Sabatier_1913, Medford_36_2015} and provides a rational basis for \ce{S} poisoning on late-TM sites.

The $E_\text{ads}$ decomposition into $\Delta E_\text{int}$ and $\Delta E_\text{dis}^{\ce{TM13}}$ contributions reveals that the interaction term magnitude dominates the total binding strength for all TMs, confirming that chemisorption is the primary driving force for \ce{S} adsorption. In contrast, the distortion contribution is generally moderate, which indicates that the ICO motif is preserved upon adsorption. However, several systems (\ce{Cr}, \ce{Mo}, \ce{Tc}, \ce{W}, \ce{Re}, \ce{Os}, and \ce{Ir}) exhibit relatively large distortion energy magnitudes (exceeding \SI{2}{\electronvolt}), indicating substantial local NC rearrangements. Importantly, the sign of $\Delta E_\text{dis}^{\ce{TM13}}$ carries a distinct physical meaning, i.e., positive values correspond to non-additive energetic penalties associated with unfavorable deformation of the pristine ICO structure, whereas negative values indicate cooperative rearrangements that lower the total energy and enhance stability. Consequently, systems with large-magnitude distortion energies combine strong chemical interactions with appreciable geometric adaptation, suggesting that \ce{S}-rich environments may locally compromise the integrity of pristine active sites even when adsorption remains exothermic.

Figure~\ref{fig3}(b) presents the evolution of the ECN and $d_{\text{av}}$ across the three TM series before and after \ce{S} adsorption. These structural descriptors provide complementary insight into adsorption-induced distortions. The $d_\text{av}$ and ECN variations, together with their respective changes ($\Delta d_\text{av}$ and $\Delta$ECN; Figures S2 and S3), indicate that most NCs retain near-ICO symmetry, with distortions localized around the adsorption site. Early \num{3}$d$ elements display larger perturbations, reflecting their lower structural rigidity and higher flexibility. In contrast, mid- and late-series NCs, particularly in the \num{4}$d$ and \num{5}$d$ rows, are more resilient, undergoing only minor geometric adjustments. Among the two descriptors, $d_\text{av}$ correlates more consistently with the distortion energy, whereas ECN exhibits heightened sensitivity to adsorption-induced rearrangements. Large ECN variations are associated with substantial structural transformations, sometimes leading to motifs more stable than the pristine ICO geometry (e.g., \ce{Cr}, \ce{Mo}, \ce{W}, and \ce{Re}), as evidenced by negative $\Delta E_\text{dis}^{\ce{TM13}}$. Conversely, positive distortion energies reflect cases in which \ce{S} adsorption acts as a perturbation that distorts the ICO framework without conferring additional stabilization. These observations demonstrate that correlating distortion energies with structural descriptors is nontrivial and requires consideration of electronic effects, including adsorption-induced changes in the total magnetic moment (Table S5).

The \ce{TM}-\ce{S} bond distances shown in Figure~\ref{fig3}(c) exhibit an approximately parabolic trend along each TM row, with the shortest distances occurring for mid-row elements, where the $d$ band is neither too empty nor too filled, thereby maximizing \ce{S}-\ce{TM} hybridization. Across a given group, the bond length generally increases from \num{3}$d$ to \num{5}$d$ elements due to atomic-size effects, although relativistic contraction in the \num{5}$d$ series leads to shorter-than-expected bonds for \ce{Ta} and \ce{Ir}. While a direct correlation between shorter \ce{TM}-\ce{S} distances and stronger adsorption might be anticipated, such a relationship is not universally observed. Structural distortions and preferential site changes, as in the case of \ce{S{\text{/}}Ir13}, generate distinct local chemical environments that modulate bond lengths independently of adsorption energy. Indeed, an inverse correlation between $d_{\ce{TM}-\ce{S}}$ and $|\Delta E_\text{int}|$ is observed for most systems, underscoring the structural origin of interaction strength. Excessively short \ce{TM}-\ce{S} bonds impose strain on the ICO cage, linking geometric frustration with the energetic fingerprints of \ce{S} adsorption.

Figure~\ref{fig3}(d) compares the $\varepsilon_d$ for pristine and \ce{S}-adsorbed NCs. The adsorption of a single \ce{S} atom induces only modest shifts in $\varepsilon_d$, indicating that the overall electronic stratification of the NCs remains largely intact and that the fundamental premises of the $d$-band model remain applicable. Across the \num{3}$d$ to \num{5}$d$ series, $\varepsilon_d$ shifts progressively to lower energies, reflecting increased $d$-states stabilization. According to the $d$-band model,\cite{Hammer_71_2000} metals with $\varepsilon_d$ closer to the Fermi level bind adsorbates more strongly, consistent with the enhanced reactivity of early TMs and the weaker \ce{S} interactions observed for late-series elements. Thus, NCs at the beginning of each TM series are the most favorable candidates for \ce{S} adsorption. From the Bader charge analysis, we observe confirmation of these electronic trends, revealing a net electron transfer from the NC to \ce{S} in all cases, driven by electronegativity differences. The largest charge-transfer occurs in \num{3}$d$ NCs, where the $d$-states are less stabilized and more readily available for donation. Differential charge-density plots further confirm localized electron accumulation on \ce{S} and depletion on neighboring metal atoms, highlighting the mixed covalent--metallic character of the \ce{S}-NC bond.

\begin{figure*}[htb!]
\centering
\includegraphics[width=0.8\linewidth]{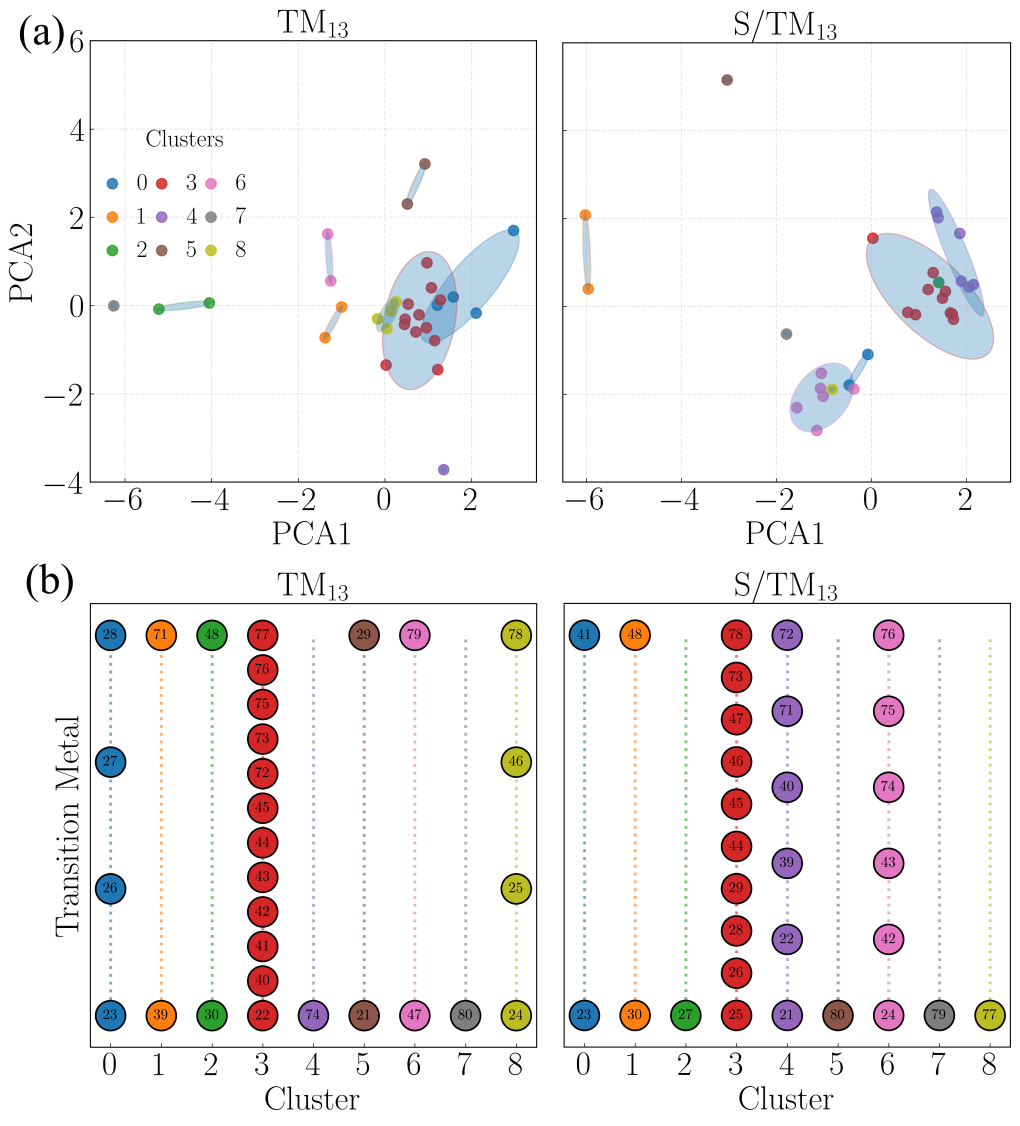}
\caption{(a) Two-dimensional PCA projections of standardized DFT-derived descriptors for pristine \ce{TM13} (left) and \ce{S{\text{/}}TM$_{13}$} (right) NCs. Colors denote \textit{k}-means cluster assignments, while ellipses indicate the dispersion of each cluster in descriptor space. (b) Cluster membership of TM-NCs obtained from \textit{k}-means classification for \ce{TM13} (left) and \ce{S{\text{/}}TM$_{13}$} (right) systems. Each point represents a TM element, identified by its atomic number, and elements that remain grouped in both classifications exhibit similar responses to \ce{S} adsorption.}
\label{fig4}
\end{figure*}

\subsection{Physics-Informed Interpretation of Sulfur Adsorption and Poisoning Trends}

\begin{figure*}[htb!]
\centering
\includegraphics[width=1.0\linewidth]{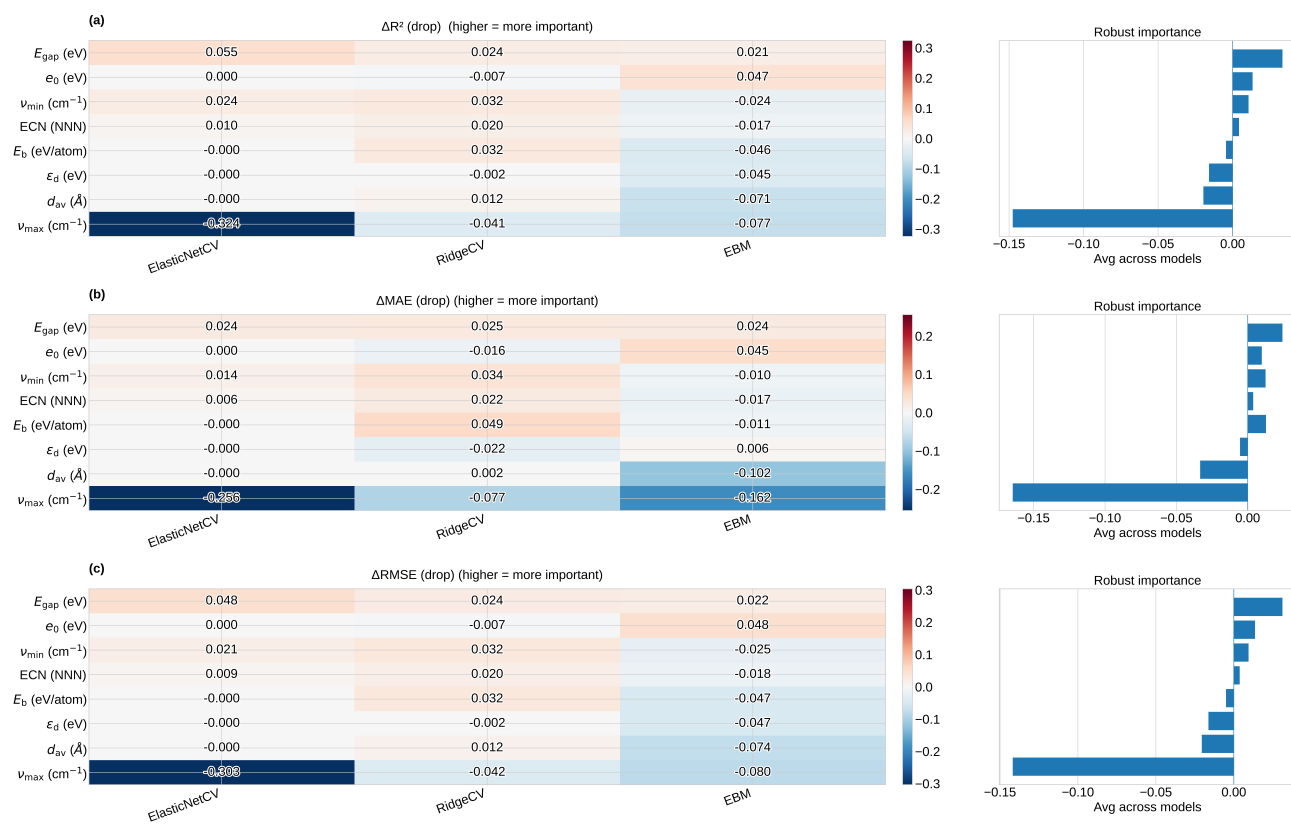}
\caption{ Leave-one-feature-out (LOFO) analysis measures how removing each pristine nanocluster descriptor affects model generalization when predicting atomic sulfur adsorption energies on \ce{TM13} icosahedra. The panels show performance drops as (a) $\Delta R^{2}$, (b) $\Delta$MAE, and (c) $\Delta$RMSE, all evaluated with group cross-validation using three regression models: ElasticNetCV, RidgeCV, and Explainable Boosting Machine (EBM). The heatmaps on the left display effects for each model, while the bar plots on the right show the average impact across models, ranking feature importance (y-axis is the same as in the left panels).}
\label{fig5}
\end{figure*}

\begin{figure*}[htb!]
\centering
\includegraphics[width=\linewidth]{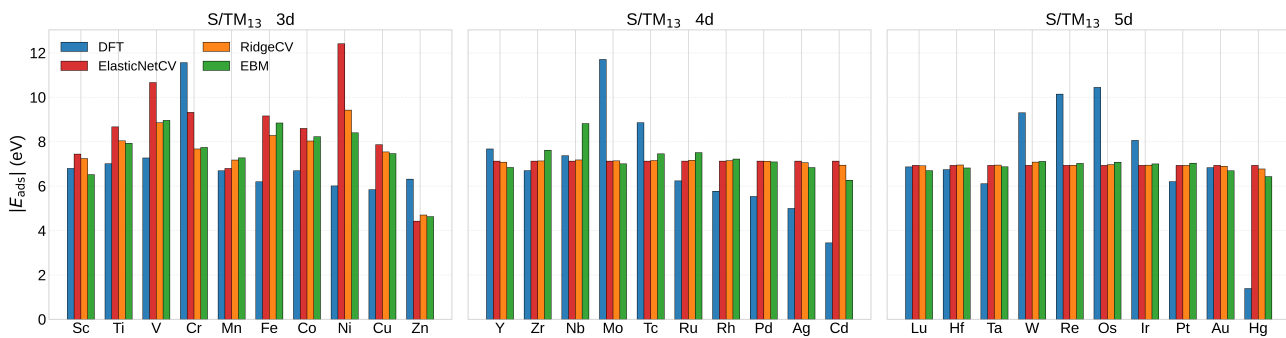}
\caption{Comparison between DFT adsorption energies with machine learning predictions for atomic sulfur adsorption on \ce{TM13} icosahedral nanoclusters across the \num{3}$d$, \num{4}$d$, and \num{5}$d$ transition metal series. For each metal, the bars show both the DFT reference value of $|E_{\mathrm{ads}}|$ for \ce{S{\text{/}}TM$_{13}$} and the predictions from three different machine learning models (ElasticNetCV, RidgeCV, and Explainable Boosting Machine, EBM) trained on the properties of the TM series.}
\label{fig6}
\end{figure*}

Here, we combine unsupervised clustering and LOFO analysis to identify periodic-table-level trends in atomic \ce{S} adsorption and to motivate our choice of representative systems (Figures~\ref{fig4}-\ref{fig6}). In descriptor space, \ce{Ti13}, \ce{Zr13}, and \ce{Hf13} (atomic numbers 22, 40, and 72) form a clear, physically meaningful group. Consistent with the DFT energetics and structural analyses discussed above, these nanoclusters exhibit strong \ce{S} binding while remaining comparatively structurally resilient within the ICO motif. Notably, \ce{Ti}, \ce{Zr}, and \ce{Hf} constitute an isoelectronic triad across the \num{3}$d$, \num{4}$d$, and \num{5}$d$ series, providing a controlled set of systems for comparing periodic trends and relativistic effects.

Figure~\ref{fig4}(a) gives a snapshot of how the descriptor space is arranged using two-dimensional PCA projections for both pristine \ce{TM13} and \ce{S{\text{/}}TM$_{13}$} systems. The way these systems split into separate regions suggests that the energetic, structural, vibrational, and electronic descriptors capture the main factors that affect \ce{S} adsorption across the transition metals. When \ce{S} adsorbs, the clusters shift in PCA space, indicating that metal-\ce{S} bonding, charge transfer, and structural changes all contribute. Still, a handful of elements end up in nearly the same spot whether they're pristine or have adsorbed \ce{S}. For example, the \ce{Ti}/\ce{Zr}/\ce{Hf} group sticks together in both the PCA projections and cluster assignments (see Figure~\ref{fig4}(b)). This stands in contrast to late transition metals, which usually bind \ce{S} more weakly, and to early or mid-series elements, which can bind strongly but tend to undergo bigger structural changes.

Considering the three regression models together with the clustering results (Figure~\ref{fig4}) and the LOFO ranking (Figure~\ref{fig5}), \ce{Ti13}, \ce{Zr13}, and \ce{Hf13} emerge in an intermediate adsorption regime: \ce{S} binds strongly enough to promote activation of \ce{S}-containing molecules, yet not so strongly that it is systematically associated with large adsorption-induced destabilization of the scaffold. This regime aligns with the Sabatier principle, which anticipates optimal catalytic behavior at intermediate binding strengths (neither too weak for activation nor too strong, leading to poisoning and slow turnover). 

Bringing together the results from both unsupervised grouping and supervised trend-finding gives us a solid, data-driven reason to pick \ce{Ti13}, \ce{Zr13}, and \ce{Hf13} as reliable and chemically tough platforms for detailed \ce{SO2} adsorption studies and related \ce{S}-poisoning situations. When prioritizing accurate predictions over the identification of general trends, the next step would be to add adsorption-response descriptors (like $\Delta E_{\mathrm{int}}$, $\Delta E_{\mathrm{dis}}$, $\Delta d_{\mathrm{av}}$, $\Delta$ECN, $d_{\mathrm{TM}-\mathrm{S}}$, or charge transfer). These will directly capture the structural and electronic changes that drive unusually strong or weak binding cases.

Our combined unsupervised and supervised ML approaches aren’t meant to replace first-principles calculations. Instead, they give us a practical, physics-informed way to understand periodic trends, see which pristine descriptors still matter during cross-validation, and pick the right materials. In this light, \ce{Ti13}, \ce{Zr13}, and \ce{Hf13} consistently stand out as a group that binds just right: \ce{S} sticks well enough to activate \ce{S}-containing molecules, but not so much that it causes significant destabilization. That sort of balance is precisely in line with the Sabatier principle. It provides a solid, data-driven rationale for choosing these nanoclusters as robust, representative platforms for detailed \ce{SO2} adsorption studies and other \ce{S}-poisoning scenarios.

\subsubsection{Case Study Validation: \ce{SO2} Adsorption}

\begin{figure*}[htb!]
\centering
\includegraphics[width=\linewidth]{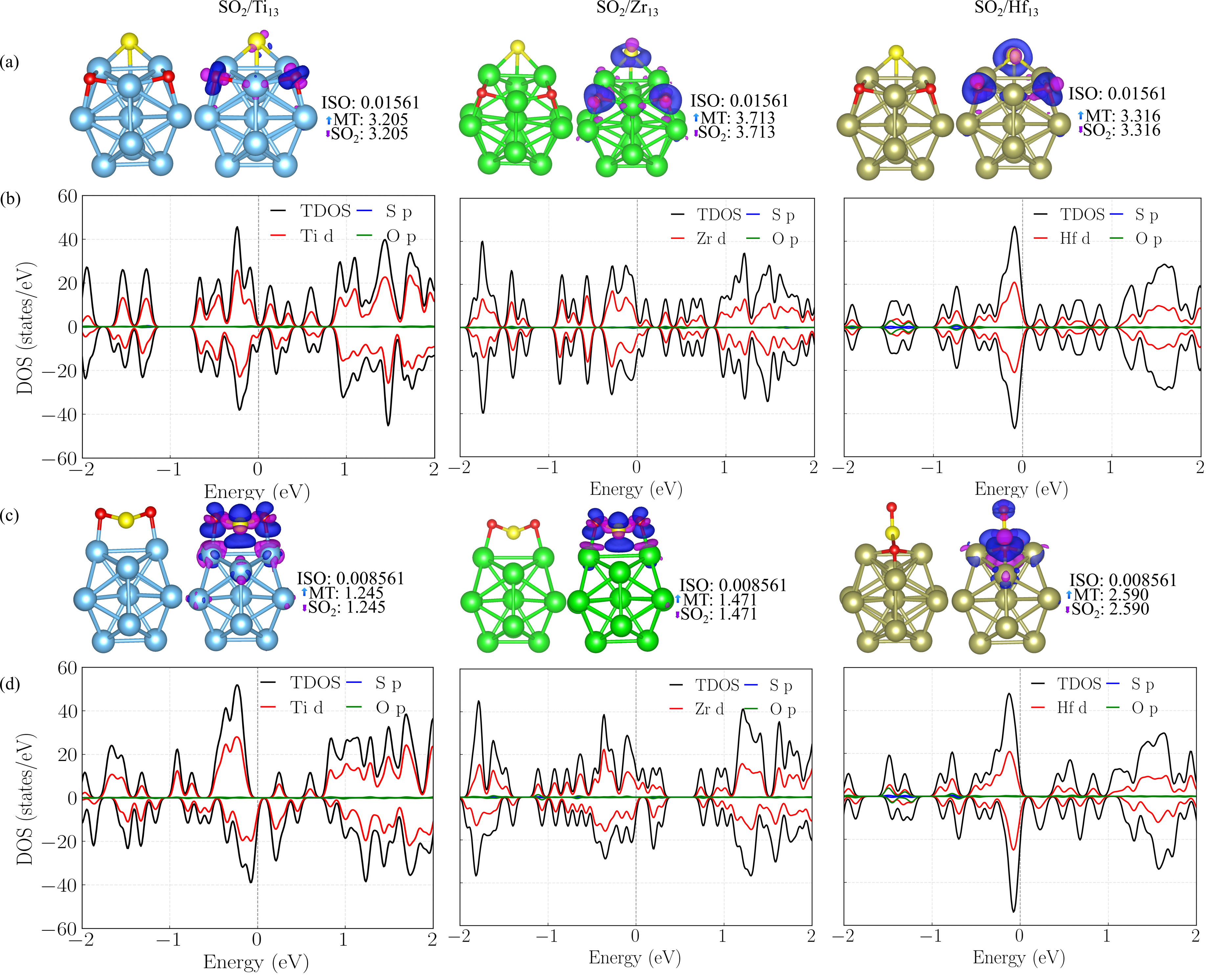}
\caption{The lowest-energy adsorption configurations and electronic structure of \ce{SO2} on \num{13}-atom ICO NCs. (a) Dissociative adsorption geometries of \ce{SO2} on \ce{Ti13}, \ce{Zr13}, and \ce{Hf13}, together with the corresponding Bader charge density difference isosurfaces. (b) Projected density of states (PDOS) for the dissociated configurations, showing the total DOS and the contributions from TM $d$, \ce{S} $p$, and \ce{O} $p$ states. (c) Most stable molecular (non-dissociated) adsorption geometries and associated charge density distributions. (d) PDOS for the intact molecular adsorption configurations. The Fermi level is set to zero energy in all panels.}
\label{fig7}
\end{figure*}

We performed explicit first-principles calculations of \ce{SO2} adsorption on the corresponding selected \num{13}-atom ICO NCs: \ce{SO$_2${\text{/}}Ti$_{13}$}, \ce{SO$_2${\text{/}}Zr$_{13}$}, and \ce{SO$_2${\text{/}}Hf$_{13}$}, using our DFT-PBE+D3 protocol, to validate the trends inferred from the ML analysis. These NCs were identified by the unsupervised learning stage as representative and chemically resilient candidates within the TM series. For each system, we have run AIMD simulations, as presented in \SM (Figure S4), exploring multiple adsorption geometries (downward \ce{S} and \ce{O} orientations and a tilted orientation) at the hollow, bridge, and top sites. In Figure~\ref{fig7}, the lowest energy dissociative adsorption geometries in panel (a) with the respective Bader isosurfaces, and the corresponding PDOS in panel (b), while the lowest energy molecular adsorption geometries are presented in panel (c) with the respective Bader isosurfaces, and the corresponding PDOS in panel (d). Additionally, in Table~\ref{tab1} are presented the $E_{\text{b}}$, $E_{\text{ads}}$, $\Delta E_{\text{int}}$, $\Delta E_{\text{dis}}$, $\Delta d_{\text{av}}$), $\Delta$ECN, $d_{\ce{TM}-\ce{S}}$, $d_{\ce{TM}-\ce{O}}$, $d_{\ce{S}-\ce{O}}$, $\varepsilon_d$, and $\nu$ values for \ce{SO$_2${\text{/}}Ti$_{13}$}, \ce{SO$_2${\text{/}}Zr$_{13}$}, and \ce{SO$_2${\text{/}}Hf$_{13}$}.

For all three NCs (\ce{Ti13}, \ce{Zr13}, and \ce{Hf13}), the putative global minimum corresponds to dissociative adsorption of the \ce{SO2} molecule. In these configurations, the \ce{S}--\ce{O} bonds are strongly elongated and effectively cleaved, leading to the formation of robust metal--\ce{S} and metal--\ce{O} bonds, as illustrated in Figure~\ref{fig7}(a). In contrast, molecularly adsorbed configurations, in which the \ce{SO2} molecule remains intact (Figure~\ref{fig7}(c)), are systematically higher in energy and therefore metastable, as quantitatively reflected by the adsorption energies listed in Table~\ref{tab1}.

\begin{table*}[t!]
\centering
\caption{The structural, energetic, electronic, and vibrational properties of the most stable \ce{SO2}-adsorbed configurations on \ce{Ti13}, \ce{Zr13}, and \ce{Hf13} ICO NCs. Values reported outside parentheses correspond to the lowest-energy dissociative adsorption configurations, whereas values in parentheses refer to the most stable molecularly adsorbed (intact) \ce{SO2} geometries. The listed quantities include the binding energy per atom ($E_{\text{b}}$), adsorption energy ($E_{\text{ads}}$), interaction and distortion energy contributions ($\Delta E_{\text{int}}$, $\Delta E_{\text{dis}}$), relative changes in average bond length ($\Delta d_{\text{av}}$) and effective coordination number ($\Delta$ECN), characteristic metal--\ce{S} and metal--\ce{O} bond lengths, the \ce{S}--\ce{O} bond distance, the $d$-band center ($\varepsilon_d$), and representative molecule vibrational frequencies ($\nu$).}
\label{tab1}
\begin{tabular}{lcccc} \toprule
 & \ce{SO$_2${\text{/}}Ti$_{13}$} & \ce{SO$_2${\text{/}}Zr$_{13}$} & \ce{SO$_2${\text{/}}Hf$_{13}$} \\ \midrule
$E_{\text{b}}$ (\si{\electronvolt/atom})                      & \num{-4.584}  (\num{-4.065}) & \num{-5.374}  (\num{-4.863}) & \num{-5.320}  (\num{-4.954}) \\
$E_\text{ads}$ (\si{\electronvolt})                              & \num{-13.063} (\num{-4.745}) & \num{-12.356} (\num{-4.179}) & \num{-12.681} (\num{-6.826}) \\
$\Delta E_{\text{int}}$ (\si{\electronvolt})                     & \num{-25.416} (\num{-6.472}) & \num{-25.41} (\num{-6.472}) & \num{-25.100} (\num{-13.182}) \\
$\Delta E_{\text{dis}}^{\ce{TM13}}$ (\si{\electronvolt/atom}) & \num{0.016} (\num{0.011}) & \num{0.360} (\num{0.043}) & \num{0.342} (\num{.288}) \\
$\Delta E_{\text{dis}}^{\ce{SO2}}$ (\si{\electronvolt/atom})  & \num{12.016} (\num{1.923}) & \num{12.591} (\num{1.733}) & \num{5.879} (\num{12.6225}) \\
$\Delta d_\text{av}$ (\SI{}{\percent})                             & \num{1.216}   (\num{0.517}) & \num{1.535}   (\num{0.497}) & \num{1.444}   (\num{1.158}) \\
$\Delta$ECN (\SI{}{\percent})                                      & \num{-1.166}  (\num{-3.595}) & \num{-1.666}  (\num{-3.017}) & \num{-1.336}  (\num{-1.633}) \\
$d_{\ce{TM}-\ce{S}}$ (\si{\angstrom})                            & \num{2.428}   (\num{2.487}) & \num{2.580}   (\num{2.670}) & \num{2.560}   (\num{2.437}) \\
$d_{\ce{TM}-\ce{O}}$ (\si{\angstrom})                            & \num{1.956}   (\num{1.913}) & \num{2.098}   (\num{2.073}) & \num{2.069}   (\num{2.068}) \\
$d_{\ce{S}-\ce{O}}$ (\si{\angstrom})                             & \num{4.198}   (\num{1.660}) & \num{4.489}   (\num{1.644}) & \num{4.410}   (\num{1.537}) \\
$\varepsilon_d$ (\si{\electronvolt})                             & \num{-0.870}   (\num{-1.125}) & \num{-1.317}   (\num{-1.345}) & \num{-1.178}   (\num{-1.181}) \\
 & \num{664.102} \; (\num{476.699}) & \num{660.040} \; (\num{446.927}) & \num{942.482} \; (\num{471.810}) \\ $\nu$ (\SI{}{\cm}$^{-1}$) &\ \num{638.769} \; (\num{438.667}) & \num{637.756} \; (\num{414.553}) & \num{540.731} \; (\num{433.483}) \\ &\ \num{464.116} \; (\num{420.727}) & \num{482.428} \; (\num{411.623}) & \num{404.675} \; (\num{420.523}) \\ \bottomrule
\end{tabular}
\end{table*}

The energetic decomposition analysis confirms that dissociative adsorption is driven by very large and negative interaction energies ($\Delta E_{\text{int}}$), which overwhelmingly dominate over the distortion penalties associated with both the NC ($\Delta E_{\text{dis}}^{\ce{TM13}}$) and the adsorbate ($\Delta E_{\text{dis}}^{\ce{SO2}}$). This behavior is particularly pronounced for \ce{Zr13} and \ce{Hf13}, which exhibit the most negative adsorption energies and interaction terms, reflecting their higher chemical affinity toward \ce{S}- and \ce{O}-containing species. Despite strong adsorption, the relatively modest $\Delta d_{\text{av}}$ and $\Delta$ECN values further indicate that the underlying ICO framework of the NCs remains largely intact, consistent with their classification as structurally resilient adsorption platforms.

Additional insight into the bonding mechanism is obtained from the PDOS shown in Figure~\ref{fig7}(b). In the dissociated configurations, the \ce{S} $p$ states hybridize strongly with the TM $d$ manifold near the Fermi level, providing a clear electronic signature of covalent metal--\ce{S} bonding. This hybridization is accompanied by a redistribution of the metal $d$ states, consistent with the downward shift of the $d$-band center reported in Table~\ref{tab1}. Such electronic reorganization is well known to stabilize strongly bound adsorbates and directly supports the trends predicted by the ML descriptors.

The structural fingerprints of dissociative adsorption are further corroborated by the pronounced elongation of the \ce{S}--\ce{O} bond distances, which increase from typical gas-phase values to more than \SI{4}{\angstrom} in the most stable configurations. This bond cleavage is also reflected in the vibrational analysis, where the disappearance of characteristic molecular \ce{SO2} stretching modes and the emergence of lower-frequency metal--\ce{S} and metal--\ce{O} vibrations provide an unambiguous spectroscopic signature of dissociation. In contrast, the preserved molecular configurations exhibit vibrational frequencies closer to gas-phase \ce{SO2}, in agreement with the structural parameters listed in parentheses in Table~\ref{tab1}. Overall, the explicit \ce{SO2} adsorption results place \ce{Ti13}, \ce{Zr13}, and \ce{Hf13} in an intermediate adsorption regime where \ce{S}-containing species bind strongly enough to undergo activation and dissociation, yet without inducing catastrophic structural NC degradation. This behavior is fully consistent with the Sabatier principle and provides direct validation of the ML-guided selection of these NCs as chemically resilient and catalytically relevant platforms for \ce{S}-containing molecules.

\section{Conclusions}

We performed a systematic investigation of the stability, reactivity, and sulfur tolerance of \num{13}-atom icosahedral transition-metal nanoclusters by combining dispersion-corrected first-principles calculations with interpretable machine-learning analyses. By screening the \num{3}$d$, \num{4}$d$, and \num{5}$d$ series within a fixed ICO motif, we isolated transferable structure-property relationships that disentangle intrinsic nanocluster stability from adsorption-driven changes. The DFT results show that the pristine-cluster binding energy follows an approximately parabolic dependence on atomic number within each $d$ series, with maximal stability near mid-series elements, consistent with optimal filling of bonding states in the cluster-orbital picture. Upon atomic \ce{S} adsorption, adsorption energetics are dominated by the interaction contribution. At the same time, distortion penalties are typically moderate, and the vibrational fingerprints reveal localized stiff \ce{TM}-\ce{S} modes and adsorption-induced broadening of the spectra. In the associated descriptor space, unsupervised clustering organizes the transition metals into chemically meaningful groups, and model-agnostic LOFO analyses across complementary regression models quantify which descriptors contribute robustly to out-of-sample prediction, thereby providing a physics-informed interpretation of sulfur adsorption and poisoning trends in the presence of descriptor collinearity.

Guided by this combined unsupervised-supervised framework, we identified the isoelectronic triad \ce{Ti13}, \ce{Zr13}, and \ce{Hf13} as representative, chemically resilient platforms that balance strong \ce{S} binding with limited adsorption-induced structural perturbation. Explicit DFT calculations of \ce{SO2} adsorption on these nanoclusters validate this selection: the preferred adsorption is strong and predominantly dissociative, driven by large metal-adsorbate interaction energies that outweigh the structural distortion costs. At the same time, the ICO framework remains largely preserved. Electronic-structure analysis shows pronounced hybridization between \ce{S}-$p$ and metal-$d$ states near the Fermi level together with a downward shift of the $d$-band center, providing a consistent electronic fingerprint for activation and poisoning propensity; concomitantly, the structural and vibrational signatures corroborate bond weakening/cleavage in \ce{SO2} and the emergence of metal--\ce{S}/metal--\ce{O} modes in the dissociated states.

Our results place \ce{Ti13}, \ce{Zr13}, and \ce{Hf13} in an intermediate adsorption regime where \ce{S}-containing molecules can be activated without catastrophic degradation of the catalytic scaffold, aligning with the Sabatier principle and highlighting these mid-series nanoclusters as promising candidates for sulfur-tolerant catalysis. More broadly, this study demonstrates that integrating first-principles energetics, electronic descriptors, and lattice-dynamical fingerprints with interpretable ML (including robustness checks across model classes) enables a rational, transferable mapping from periodic-table chemistry to adsorption and poisoning mechanisms in sub-nanometer catalysts.

\section*{Acknowledgements}
The project (HGF ZT-I-PF-5-261 GENIUS) underlying this publication is/was funded by the Initiative and Networking Fund of the Helmholtz Association in the framework of the Helmholtz AI project call.
The authors are thankful for financial support from the National Council for Scientific and Technological Development (CNPq, grant numbers 303206/2025-0, 444431/2024-1, 408144/2022-0, 305174/2023-1, and 444069/2024-0), the Rio Grande do Sul Research Foundation (FAPERGS, grant number 24/2551-0001551-5), Federal District Research Support Foundation (FAPDF, grants 00193-00001817/2023-43 and 00193-00002073/2023-84), PDPG-FAPDF-CAPES Centro-Oeste (grant number 00193-00000867/2024-94), and the Coordination for Improvement of Higher Level Education (CAPES). In addition, the authors thanks the ``Centro Nacional de Processamento de Alto Desempenho em S\~ao Paulo'' (CENAPAD-SP, UNICAMP/FINEP - MCTI project) for resources into the 897 and 570 projects, Lobo Carneiro HPC (NACAD) at the Federal University of Rio de Janeiro (UFRJ) for resources (133 project) CIMATEC SENAI at Salvador -- BA, Brazil for the partnership, support through the Ogun Supercomputer, and ``Laborat{\'o}rio Central de Processamento de Alto Desempenho'' (LCPAD) financed by FINEP through CT-INFRA/UFPR projects and Santos Dumont/LNCC.

\section*{Author contributions statement}
M.J.P. conceived the project. M.J.P., R.L.T.P., D.G.S., C.R.C.R., and A.C.D. provided the computational resources. R.F.M., J.M.T.P., T.G.G., and M.J.P. performed the calculations. O.R.F., M.J.P., and C.R.C.R. curated the data and developed the algorithmic codes. R.F.M., J.M.T.P., T.G.G., O.R.F., K.E.A.B., and M.J.P. prepared the figures and tables and drafted the initial version of the manuscript. R.L.T.P., D.G.S., C.R.C.R., A.C.D., and K.E.A.B. contributed to data curation and critically reviewed the manuscript. M.J.P. supervised the project and led the manuscript preparation. All authors contributed to the writing and revision of the manuscript.

\section{Data availability}
In addition to the supporting information, the data sets used and/or analyzed during the current study are available on the following repository \href{URL}{https://github.com/KIT-Workflows/Physics-Informed-Nanoclusters}.

\section{Additional information}
Competing Interests: The authors declare that they have no competing interests.

\bibliography{jshort,reference}
\bibliographystyle{rsc}

\end{document}


\maketitle

\tableofcontents

\newpage

\section{Convergence Tests}

\begin{table*}[ht!]
\caption*{Table S1: Convergence tests for \ce{Hg13} with regard to the box size ({\sl{Box Size}}): the relative total energy ($\Delta E_\text{tot}$), average bond length ($d_{av}$), effective coordination number (ECN), and total magnetic moment ($m_\text{T}$).}
\label{tab1}
\centering
\scalebox{1.0}{
\begin{tabular} {lccccccc} \hline
{\sl{Box Size}} (\SI{}{\angstrom}) & $\Delta E_\text{tot}$ (\SI{}{\electronvolt}) & ECN & $d_{av}$ (\SI{}{\angstrom}) & $m_\text{T}$ (\si{\micro_B}) \\ \hline
$12$ & $0.1910$ & $6.3719$ & $2.9493$ & $6.0000$ \\
$14$ & $0.0088$ & $6.3722$ & $2.9461$ & $6.0000$ \\		
$16$ & $0.0062$ & $6.3940$ & $2.9457$ & $6.0000$ \\ 		
$18$ & $0.0002$ & $6.3914$ & $2.9454$ & $6.0000$ \\ 		
$20$ & $0.0001$ & $6.3913$ & $2.9455$ & $6.0000$ \\ 		
$22$ & $0.0000$ & $6.3914$ & $2.9455$ & $6.0000$ \\  \hline
\end{tabular}
}
\end{table*}

\begin{table*}[ht!]
\caption*{Table S2: Convergence tests for \ce{Hg13} with regard to the cutoff energy ({\sl{ENCUT}}): the relative total energy ($\Delta E_\text{tot}$), average bond length ($d_{av}$), effective coordination number (ECN), and total magnetic moment ($m_\text{T}$).}
\label{tab2}
\centering
\scalebox{1.0}{
\begin{tabular} {lccccccc} \hline
{\sl{ENCUT}} (\SI{}{\electronvolt}) & $\Delta E_\text{tot}$ (\SI{}{\electronvolt}) & ECN & $d_{av}$ (\SI{}{\angstrom}) & $m_\text{T}$ (\si{\micro_B}) \\ \hline
$ 70.7410$ & $4.1910$ & $6.3709$ & $2.9552$ & $4.0000$ \\
$141.4820$ & $0.3288$ & $6.3725$ & $2.9472$ & $2.0000$ \\		
$282.9640$ & $0.0962$ & $6.3941$ & $2.9459$ & $6.0000$ \\ 		
$424.4460$ & $0.0023$ & $6.3913$ & $2.9455$ & $6.0000$ \\ 		
$495.1870$ & $0.0011$ & $6.3914$ & $2.9454$ & $6.0000$ \\ 		
$565.9280$ & $0.0000$ & $6.3914$ & $2.9455$ & $6.0000$ \\  \hline
\end{tabular}
}
\end{table*}

\begin{table*}[ht!]
\caption*{Table S3: Convergence tests for \ce{Hg13} with regard to the energy criterion (electronic convergence, {\sl{EDIFF}}): the relative total energy ($\Delta E_\text{tot}$), average bond length ($d_{av}$), effective coordination number (ECN), and total magnetic moment ($m_\text{T}$).}
\label{tab3}
\centering
\scalebox{1.0}{
\begin{tabular} {lccccccc} \hline
{\sl{EDIFF}} (\SI{}{\electronvolt}) & $\Delta E_\text{tot}$ (\SI{}{\electronvolt}) & ECN & $d_{av}$ (\SI{}{\angstrom}) & $m_\text{T}$ (\si{\micro_B}) \\ \hline
$10^{-2}$ & $0.1910$ & $6.3799$ & $2.9498$ & $6.0000$ \\
$10^{-3}$ & $0.0097$ & $6.3865$ & $2.9459$ & $6.0000$ \\		
$10^{-4}$ & $0.0034$ & $6.3913$ & $2.9454$ & $6.0000$ \\ 		
$10^{-5}$ & $0.0002$ & $6.3914$ & $2.9455$ & $6.0000$ \\ 		
$10^{-6}$ & $0.0001$ & $6.3914$ & $2.9455$ & $6.0000$ \\ 		
$10^{-7}$ & $0.0000$ & $6.3914$ & $2.9455$ & $6.0000$ \\  \hline
\end{tabular}
}
\end{table*}

\begin{table*}[ht!]
\caption*{Table S4: Convergence tests for \ce{Hg13} with regard to the force criterion (ionic convergence, {\sl{EDIFFG}}): the relative total energy ($\Delta E_\text{tot}$), average bond length ($d_{av}$), effective coordination number (ECN), and total magnetic moment ($m_\text{T}$).}
\label{tab4}
\centering
\scalebox{1.0}{
\begin{tabular} {lccccccc} \hline
{\sl{EDIFF}} (\SI{}{{\electronvolt/}{\angstrom}}) & $\Delta E_\text{tot}$ (\SI{}{\electronvolt}) & ECN & $d_{av}$ (\SI{}{\angstrom}) & $m_\text{T}$ (\si{\micro_B}) \\ \hline
$0.100$ & $0.1787$ & $6.3701$ & $2.9503$ & $6.0000$ \\
$0.050$ & $0.0113$ & $6.3987$ & $2.9469$ & $6.0000$ \\		
$0.025$ & $0.0024$ & $6.3912$ & $2.9456$ & $6.0000$ \\ 		
$0.015$ & $0.0002$ & $6.3913$ & $2.9455$ & $6.0000$ \\ 		
$0.010$ & $0.0002$ & $6.3913$ & $2.9455$ & $6.0000$ \\ 		
$0.005$ & $0.0000$ & $6.3914$ & $2.9454$ & $6.0000$ \\  \hline
\end{tabular}
}
\end{table*}

\section{Lowest energy \ce{TM13} and \ce{S{\text{/}}TM$_{13}$}}

\begin{figure}[H]
    \centering
    \includegraphics[width=1\linewidth]{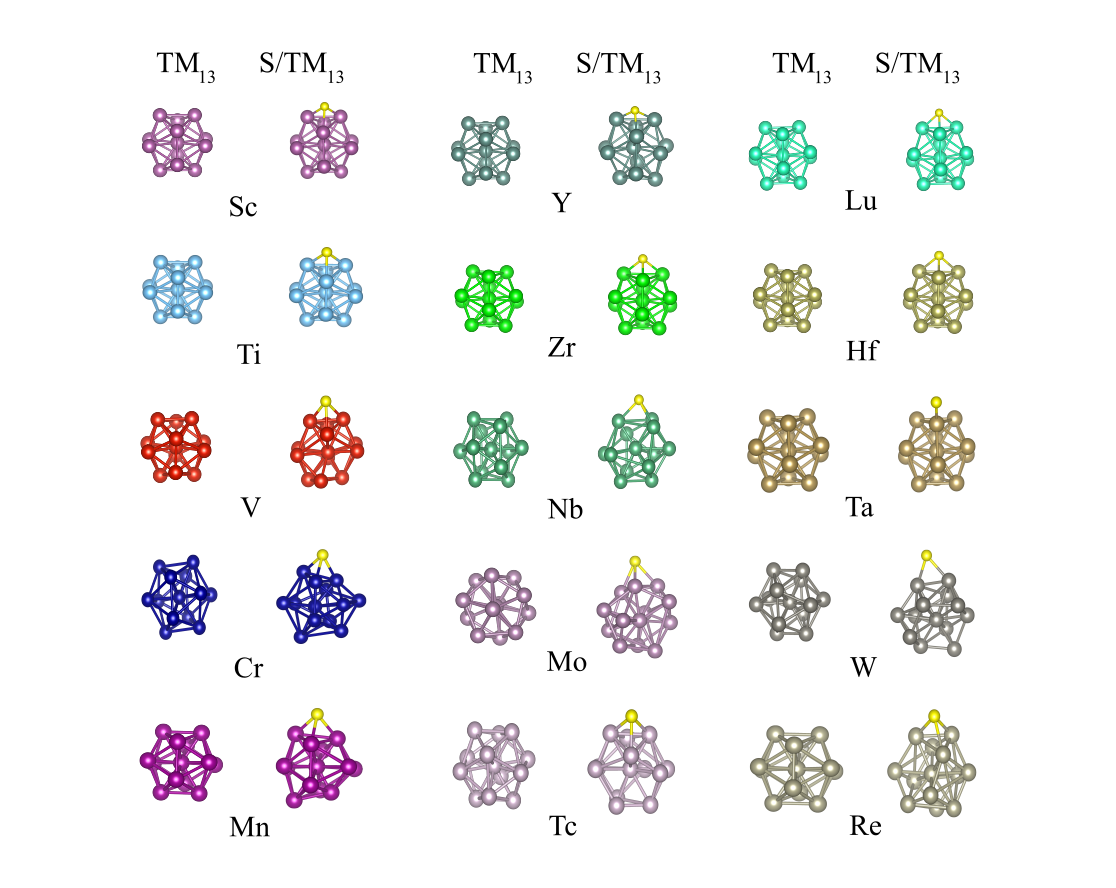}
    \caption*{Figure S1 (first part): Lowest energy \ce{TM13} and \ce{S{\text{/}}TM$_{13}$} systems.}
    \label{figs1}
\end{figure}

\begin{figure}[H]
    \centering
    \includegraphics[width=1\linewidth]{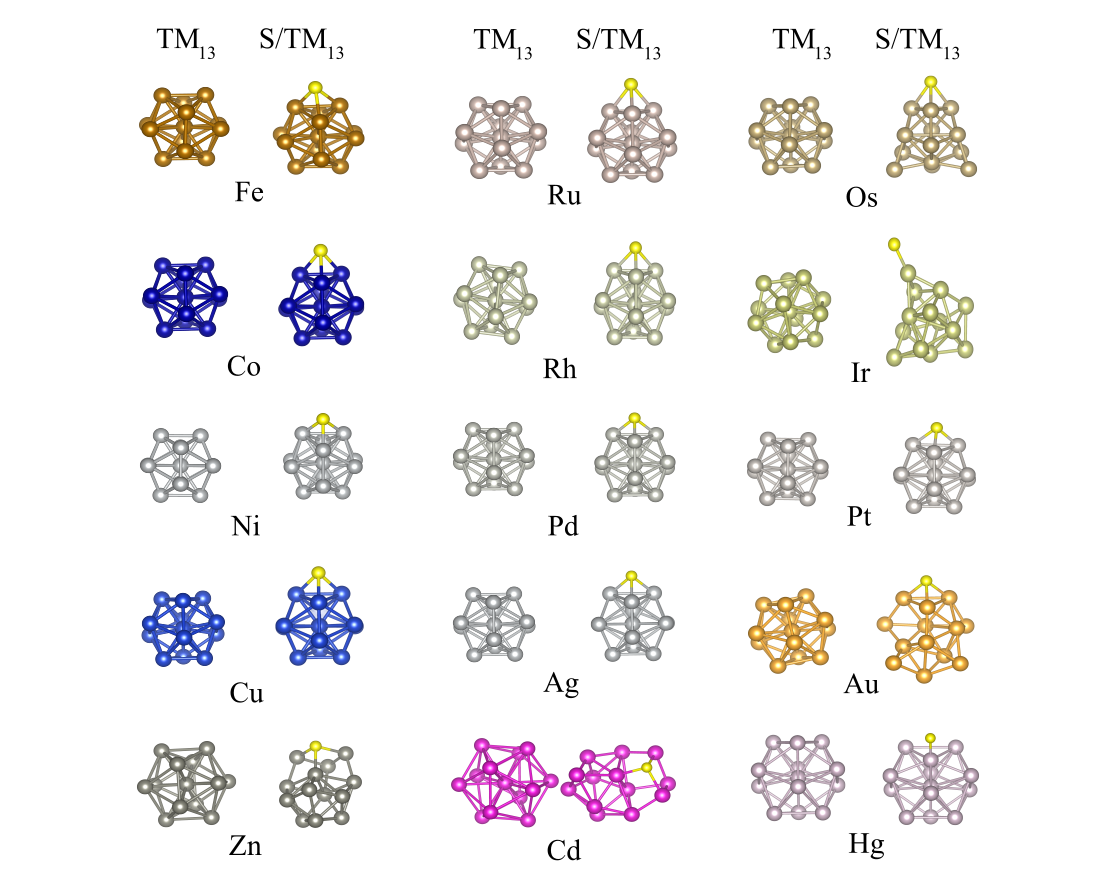}
    \caption*{Figure S1 (second part): Lowest energy \ce{TM13} and \ce{S{\text{/}}TM$_{13}$} systems.}
    \label{figs1}
\end{figure}

\subsection{\ce{TM13} Properties}

\begin{figure}[H]
    \centering
    \includegraphics[width=1\linewidth]{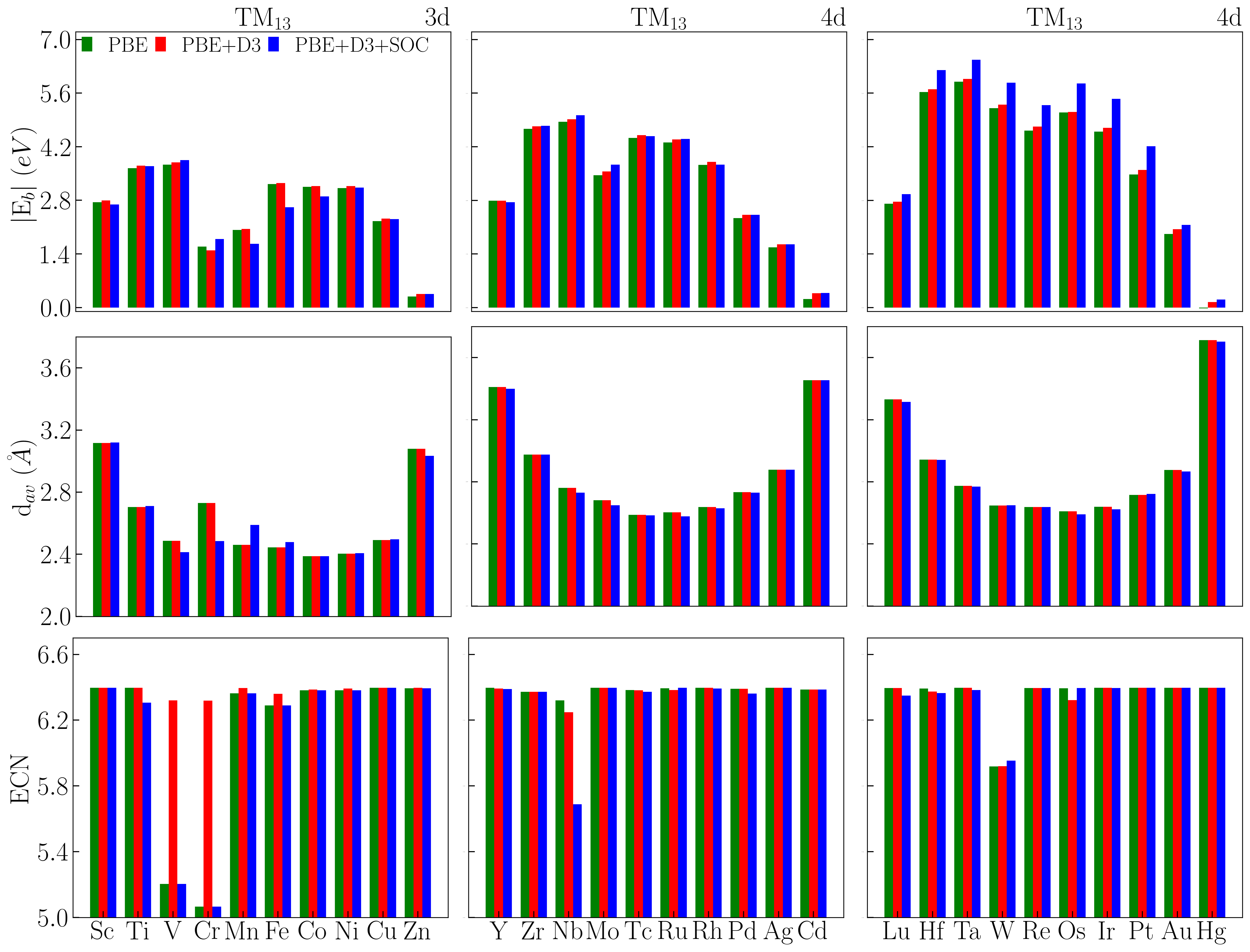}
    \caption*{Figure S2: \ce{TM13} properties: Magnitude of binding energy per atom ($|E_\text{b}|$), average bond length ($d_{\text{av}}$), and the effective coordination number (ECN) as a function of atomic number. Considering the PBE, PBE+D3, and PBE+D3+SOC calculation protocols.}
    \label{figs2}
\end{figure}

\subsection{\ce{S{\text{/}}TM$_{13}$} Properties}

\begin{figure}[H]
    \centering
    \includegraphics[width=1.0\linewidth]{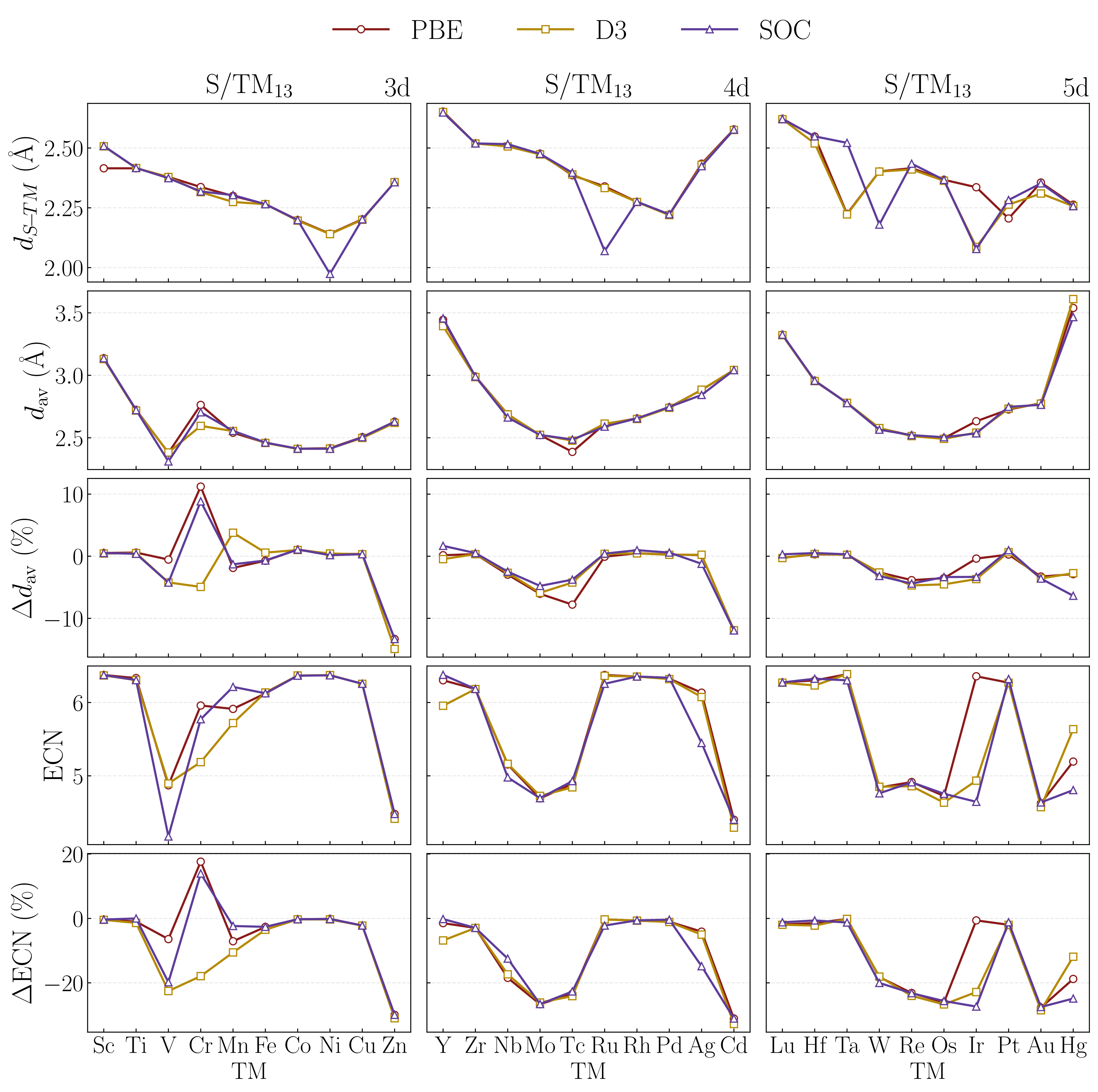}
    \caption*{Figure S3: \ce{S{\text{/}}TM$_{13}$} properties: Average distance between \ce{S} and TM atoms ($d_{S-TM}$), average bond length ($d_\text{av}$), the relative variation for $d_\text{av}$, effective coordination number (ECN), and  the relative variation for ECN as a function of atomic number. Considering the PBE, D3, and SOC calculation protocols.}
    \label{figs3}
\end{figure}

\section{AIMD -- Thermalization}

\begin{figure}[H]
    \centering
    \includegraphics[width=0.50\linewidth]{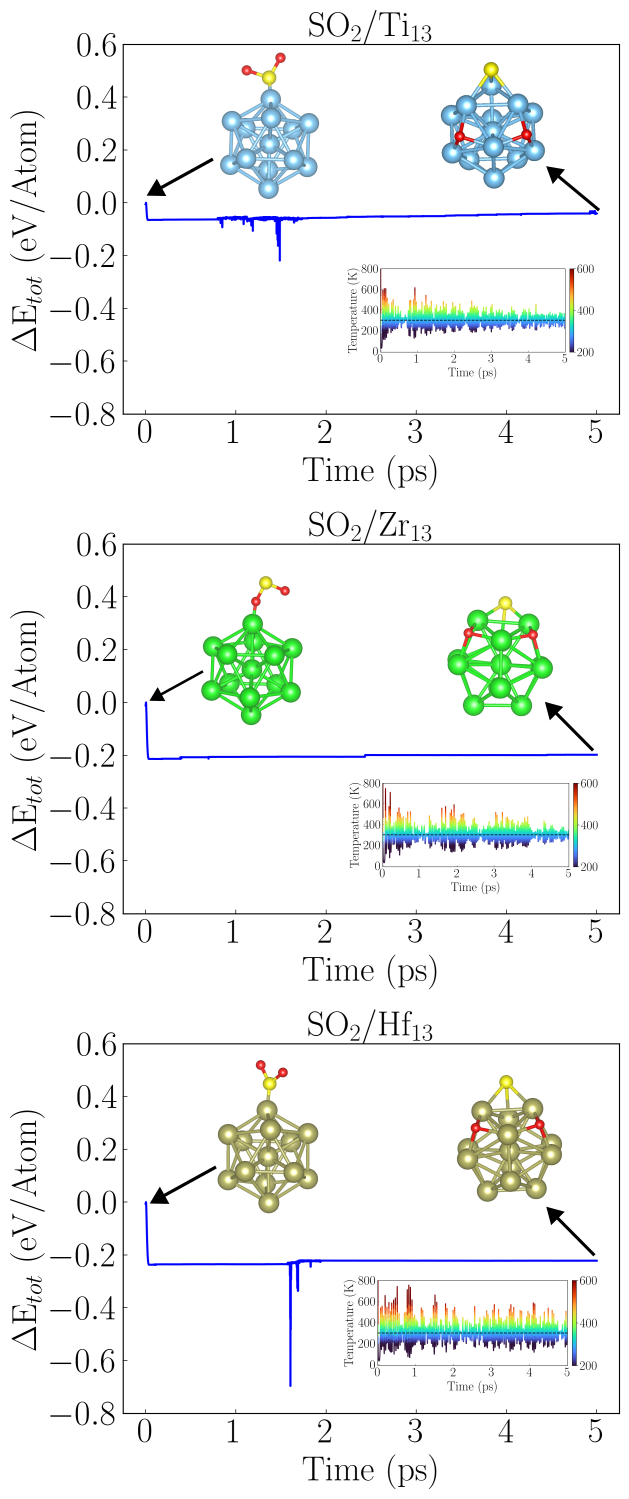}
    \caption*{Figure S4: The \textit{Ab~Initio} Molecular Dynamics (AIMD), considering thermalization protocol, for \ce{SO$_2${\text{/}}Ti$_{13}$}, \ce{SO$_2${\text{/}}Zr$_{13}$}, and \ce{SO$_2${\text{/}}Hf$_{13}$} systems.}
    \label{figs4}
\end{figure}

\newpage
\section{Total Magnetic Moment Values}

\begin{table*}[h!]
\centering
\caption*{Table S5: Total magnetic moment values, $m_\text{T}$ (\si{\micro_B}) for \ce{TM13} and \ce{S{\text{/}}TM$_{13}$} systems, for all TM elements.}

\label{tab:Mtot_TM13}
\begin{tabular}{lcc|lcc|lcc}
\hline
System & \ce{TM13} & \ce{S/TM13} &
System & \ce{TM13} & \ce{S/TM13} &
System & \ce{TM13} & \ce{S/TM13} \\
\hline
\ce{Sc13} & 19 & 17 &
\ce{Y13}  & 7  & 1  &
\ce{Lu13} & 13 & 11 \\

\ce{Ti13} & 6  & 6  &
\ce{Zr13} & 6  & 2  &
\ce{Hf13} & 6  & 4  \\

\ce{V13}  & 7  & 3  &
\ce{Nb13} & 3  & 1  &
\ce{Ta13} & 7  & 5  \\

\ce{Cr13} & 50 & 32 &
\ce{Mo13} & 20 & 2  &
\ce{W13}  & 4  & 4  \\

\ce{Mn13} & 33 & 29 &
\ce{Tc13} & 13 & 1  &
\ce{Re13} & 13 & 1  \\

\ce{Fe13} & 34 & 40 &
\ce{Ru13} & 18 & 18 &
\ce{Os13} & 16 & 2  \\

\ce{Co13} & 21 & 27 &
\ce{Rh13} & 17 & 17 &
\ce{Ir13} & 15 & 7  \\

\ce{Ni13} & 8  & 8  &
\ce{Pd13} & 8  & 4  &
\ce{Pt13} & 2  & 2  \\

\ce{Cu13} & 5  & 1  &
\ce{Ag13} & 5  & 1  &
\ce{Au13} & 5  & 1  \\

\ce{Zn13} & 0  & 0  &
\ce{Cd13} & 0  & 0  &
\ce{Hg13} & 0  & 0  \\
\hline
\end{tabular}
\end{table*}